\documentclass[prx,aps,twocolumn,preprintnumbers,amsmath, amssymb,superscriptaddress,showpacs]{revtex4-1}

\usepackage{amsmath}
\usepackage[pdftex]{graphicx}
\usepackage{amssymb}
\usepackage{color}
\usepackage[cmyk]{xcolor}
\usepackage{subfigure}
\usepackage{CJK}
\usepackage{tikz}
\usetikzlibrary{trees,snakes}

\usepackage{indentfirst}
\usepackage{amsmath}
\usepackage{cases}
\usepackage[pdftex,bookmarks=true,colorlinks=true,linkcolor=blue,urlcolor=blue,citecolor=blue]{hyperref}

\newcommand{\be}{\begin{equation}}
\newcommand{\ee}{\end{equation}}

\newcommand{\bit}{\begin{itemize}}
\newcommand{\eit}{\end{itemize}}
\newcommand{\bea}{\begin{eqnarray}}
\newcommand{\eea}{\end{eqnarray}}

\usepackage{epsfig}
\usepackage{color}
\usepackage{bbm}

\tikzset{dline/.style={color=black,line width=0.2pt}}

\begin{document}
%\linenumbers

\title
{Accurate determination of low-energy eigenspectra with multi-target matrix product states}

\author{Xuan Li}
\affiliation
{Beijing National Laboratory for Condensed Matter Physics and Institute of Physics, Chinese Academy of Sciences, Beijing 100190, China}
\affiliation
{School of Physical Sciences, University of Chinese Academy of Sciences, Beijing 100049, China}

\author{Zongsheng Zhou}
\affiliation
{Beijing National Laboratory for Condensed Matter Physics and Institute of Physics, Chinese Academy of Sciences, Beijing 100190, China}

\author{Guanglei Xu}
\affiliation
{Beijing National Laboratory for Condensed Matter Physics and Institute of Physics, Chinese Academy of Sciences, Beijing 100190, China}

\author{Runze Chi}
\affiliation
{Beijing National Laboratory for Condensed Matter Physics and Institute of Physics, Chinese Academy of Sciences, Beijing 100190, China}
\affiliation
{School of Physical Sciences, University of Chinese Academy of Sciences, Beijing 100049, China}

\author{Yibin Guo}
\affiliation
{Beijing National Laboratory for Condensed Matter Physics and Institute of Physics, Chinese Academy of Sciences, Beijing 100190, China}
\affiliation
{School of Physical Sciences, University of Chinese Academy of Sciences, Beijing 100049, China}

\author{Tong Liu}
\affiliation
{Beijing National Laboratory for Condensed Matter Physics and Institute of Physics, Chinese Academy of Sciences, Beijing 100190, China}
\affiliation
{School of Physical Sciences, University of Chinese Academy of Sciences, Beijing 100049, China}

\author{Haijun Liao}
\affiliation
{Beijing National Laboratory for Condensed Matter Physics and Institute of Physics, Chinese Academy of Sciences, Beijing 100190, China}
\affiliation
{Songshan Lake Materials Laboratory, Dongguan, Guangdong 523808, China}

\author{Tao Xiang} \email{txiang@iphy.ac.cn}
\affiliation
{Beijing National Laboratory for Condensed Matter Physics and Institute of Physics, Chinese Academy of Sciences, Beijing 100190, China}
\affiliation
{School of Physical Sciences, University of Chinese Academy of Sciences, Beijing 100049, China}
\affiliation
{Beijing Academy of Quantum Information Sciences, Beijing, 100190, China}

\date{\today}

\begin{abstract}

  Determining the low-energy eigenspectra of quantum many-body systems is a long-standing challenge in physics. In this work, we solve this problem by introducing two novel algorithms to determine low-energy eigenstates based on a compact matrix product state (MPS) representation of the multiple targeted eigenstates. The first algorithm utilizes a canonicalization approach that takes advantage of the imaginary-time evolution of multi-target MPS, offering faster convergence and ease of implementation. The second algorithm employs a variational approach that optimizes local tensors on the Grassmann manifold, capable of achieving higher accuracy. These algorithms can be used independently or combined to enhance convergence speed and accuracy. We apply them to the transverse-field Ising model and demonstrate that the calculated low-energy eigenspectra agree remarkably well with the exact solution. Moreover, the eigenenergies exhibit uniform convergence in gapped phases, suggesting that the low-energy excited eigenstates have nearly the same level of accuracy as the ground state. Our results highlight the accuracy and versatility of multi-target MPS-based algorithms for determining low-energy eigenspectra and their potential applications in quantum many-body physics.

\end{abstract}

\maketitle

\section{Introduction}

 In the study of many-body quantum systems, excited states offer more physical information than the ground state and are closely associated with various novel or unresolved physical phenomena \cite{RevModPhys.80.517, PhysRevB.48.10345, Yang_2023}. As a result, the investigation of excited states has led to the emergence of new research fields and insights into open questions. For instance, the behavior of energy eigenspectra plays a crucial role in classifying quantum phase transitions according to Ehrenfest's classification in the thermodynamic limit \cite{Jaeger1998TheEC}. A crossing between two energy levels indicates a first-order quantum phase transition involving the swap of two wave functions. In contrast, if two energy levels gradually touch each other but without crossing, it leads to a continuous quantum phase transition \cite{PhysRevLett.99.100601, PhysRevLett.91.162502}. Hence by detecting the level crossing, we can distinguish quantum phase transitions and determine the critical points \cite{PhysRevLett.104.137204, PhysRevLett.121.107202, Wang_2022}.

 Moreover, a quantum phase transition may directly arise from excited states when the vanishing gap between the ground state and the first excited state does not occur in isolation but in conjunction with the clustering of levels near the ground state. This local divergence in the density of excited states propagates to higher excitation energies as the control parameter varies, leading to an excited-state quantum phase transition. Various many-body quantum systems have been found to exhibit such transitions, including the dynamical Hamiltonian \cite{heyl2018dynamical,PhysRevLett.124.043001,PhysRevA.83.033802}, Lipkin-Meshkov-Glick model \cite{PhysRevLett.95.050402,Heiss_2005,PhysRevA.94.012113}, Dicke model \cite{PhysRevE.88.032133,PhysRevA.87.023819}, interacting boson model \cite{PhysRevA.80.032111}, kicked-top model \cite{PhysRevLett.112.140408}, vibron model \cite{PhysRevA.81.050101}, and others.

 Many-body localization systems undergo a quantum transition from the ergodic to the localized regime, which remains elusive at finite energy densities. This is attributed to the low entanglement entropy of highly excited states and the presence of numerous local excitations \cite{PhysRevLett.116.247204, PhysRevLett.118.017201, PhysRevX.10.021051}. Excited states exhibit fascinating physical properties beyond their energy levels. For example, the entanglement entropy of excited states related to the primary field exhibits universal scaling similar to the ground state in a one-dimensional critical model \cite{PhysRevLett.106.201601}. Additionally, the study of entanglement away from the critical point is also important \cite{PhysRevA.70.032313, Alba_2009}. Analysis of low-energy excited states can provide conformal data, one of the most important physical quantities used to identify the universal class of a phase transition, with unprecedented accuracy \cite{PhysRevLett.121.230402}. In some nonequilibrium systems, consideration of the low-lying excited states is necessary to avoid inaccurate results and extend the reliable predicted time \cite{PhysRevLett.91.049701}. Research on excited states is expected to find new quantum protocols and settle some outstanding problems.

  Exact diagonalization is a reliable method to obtain all eigenstates for small many-body systems. However, the computational time and cost become prohibitively large for larger systems due to the exponential growth of the Hilbert space. This limitation restricts the system size to approximately two dozen sites, leading to finite size effects that can obscure the understanding of certain physical properties \cite{PhysRevLett.104.137204, PhysRevLett.113.107204, PhysRevB.91.081103}. In contrast, the density matrix renormalization group (DMRG) provides a powerful tool for studying large quantum lattice systems \cite{PhysRevLett.69.2863}. The success of DMRG relies on the fact that the wave function generated by this method is an MPS~\cite{PhysRevLett.75.3537}, which captures faithfully the entanglement structure of the ground state.

  Although DMRG provides unprecedented accuracy in calculating ground state properties of one-dimensional systems, computing excited states can be challenging. Various schemes have been proposed to calculate low-energy excited eigenstates. If the excited states of interest are the lowest-lying states of different sectors distinguished by symmetry, they can be obtained by searching for the lowest energy states in those sectors \cite{PhysRevB.48.10345}. However, if the system lacks symmetry or the states of interest are in the same symmetry sector, new algorithms are required. One such approach is the multi-target DMRG \cite{PhysRevB.48.10345}, where multiple low-energy eigenstates are calculated by diagonalizing the renormalized Hamiltonian obtained in the DMRG sweep. This renormalized Hamiltonian is obtained by minimizing the truncation error of the reduced density matrix defined by summing over all targeted states with certain weights \cite{PhysRevB.48.10345, PhysRevLett.91.049701, RevModPhys.77.259}.

 Another way to compute excited states is to add a penalty term to the original Hamiltonian $H$, resulting in a modified Hamiltonian
 \begin{equation}
 H_{1} = H + \lambda|\psi_{0}\rangle\langle\psi_{0}|,
 \end{equation}
 where $|\psi_{0}\rangle$ is the ground state of $H$. The penalty term shifts the original lowest energy $E_{0}$ to $E_{0} + \lambda$, making the first excited state of $H$ the ground state of $H_{1}$ for sufficiently large $\lambda$. In principle, higher excitation states can be obtained by repeating the procedure, but this approach faces challenges to converge due to the high accuracy required for the states in the penalty term \cite{Bañuls2013}.

 The dynamical correlation function provides excitation spectra by evaluating time-dependent DMRG or time-evolving block decimation (TEBD) \cite{PhysRevLett.91.147902} from the ground state. However, it can be computationally expensive, and artificial extrapolation is often used to improve the frequency resolution \cite{PhysRevB.77.134437}. Recently, the tangent-space method based on the single mode approximation has been developed to capture excitations on the lattice with translation invariance \cite{Ostlund1995,Haegeman2012,Vanderstraeten2015,Vanderstraeten2019,Ponsioen2020,Ponsioen2022,Chi2022}.

 In this work, we show that the wave functions generated by the multi-target DMRG can be represented as a group of MPS that share a common set of matrices and each individual MPS corresponds to a specific target state of the system. We call this kind of wave function a multi-target MPS. Based on this multi-target MPS representation, we propose two novel algorithms to determine low-energy eigenspectra accurately and simultaneously. The first method, referred to as the multi-target update method (MTU), is similar to the conventional update \cite{PhysRevLett.101.090603} or TEBD method for the ground state \cite{PhysRevLett.91.147902}, but optimized for a batch of states simultaneously with reorthonormalization after each projection step. The virtual bond dimension is proportional to the number of target states to contain more entanglement. The second method, referred to as the variational Riemannian optimization (VRO), utilizes a subspace formed by isometric matrices to perform optimization without the normalized denominator that may diverge \cite{10.21468/SciPostPhys.10.2.040}. VRO can achieve accurate results with a small virtual bond dimension by preserving the orthonormalization of states and satisfying the Ring-Wirth nonexpansive condition \cite{RingOpti}, enabling the algorithm to be globally convergent. While MTU has faster convergence, the virtual bond dimension needs to increase with the number of states, VRO can significantly improve the numerical accuracy based on the results of MTU.

 As will be discussed, the use of multi-target MPS allows for the efficient computation of properties of multiple target states simultaneously, without the need to perform separate calculations for each state. This can be particularly useful in situations where one is interested in studying the properties of multiple low-energy states of the system.

 We test the two methods by evaluating the low-energy eigenspectra of a finite transverse-field Ising chain with open boundary conditions. The simulated results agree excellently with the exact solution. This demonstrates the reliability and potential of our proposal. The absolute errors of the eigenspectra and the variances in energy exhibit striking uniform convergence in the gapped phase, indicating that it is possible to compute numerous low-lying excited states with nearly the same precision.

\newcommand{\LeftIsometry}[6]{
 \draw (#1,#2) -- (#1+#3*0.75,#2) -- (#1+#3*1.15,#2+0.5*#4) -- (#1+#3*0.75,#2+#4) -- (#1,#2+#4) -- cycle ;
 \draw (#1,#2+#4*0.5) -- (#1-0.75, #2+#4*0.5)
       (#1+#3*1.15,#2+0.5*#4) -- (#1+#3*1.15+0.75,#2+0.5*#4) ;
 \draw (#1+#3*0.5,#2+#4) -- (#1+#3*0.5,#2+#4+0.75) node[above] {$#5$}
       (#1+#3*0.5,#2) node[below] {$#6$} ;
}

\newcommand{\RightIsometry}[6]{
 \draw (#1,#2) -- (#1-#3*0.75,#2) -- (#1-#3*1.15,#2+0.5*#4) -- (#1-#3*0.75,#2+#4) -- (#1,#2+#4) -- cycle ;
 \draw (#1,#2+#4*0.5) -- (#1+0.75, #2+#4*0.5)
       (#1-#3*1.15,#2+0.5*#4) -- (#1-#3*1.15-0.75,#2+0.5*#4) ;
 \draw (#1-#3*0.5,#2+#4) -- (#1-#3*0.5,#2+#4+0.75) node[above] {$#5$}
       (#1-#3*0.5,#2) node[below] {$#6$} ;
}

\section{MPS parametrization of the multi-target DMRG states}

 To construct the MPS representation for the eigenstates generated by the multi-target DMRG method \cite{PhysRevB.48.10345}, let us first consider how the MPS representation is obtained for the ground state obtained in a DMRG calculation \cite{PhysRevLett.69.2863, PhysRevLett.75.3537}. In the standard DMRG calculation, a system, known as a superblock, is partitioned into four parts: a left subblock, a right subblock, and two added lattice sites. Assuming $i$ and $i+1$ are the two added sites, then the left block contains all the sites on the left of $i$ and the right block contains all the sites on the right of $i+1$. If we use $|s_{i-1}\rangle$ and $|e_{i+2}\rangle$ to represent the basis states retained in the DMRG iterations for the left and right subblocks, then the ground state is  as $\psi_0 (s_{i-1} ,\sigma_i, \sigma_{i+1} ,e_{i+2})$ with $\sigma_j$ $(j=i, i+1)$ the quantum number of the basis states at site $j$.

 Now we divide the superblock into two parts, a system block plus an environment block. The system block contains the left block plus site $i$. The environment block, on the other hand, contains the right block plus site $i+1$.  In this bipartite representation, $\psi_0 $ can be regarded as a matrix with $(s_{i-1} , \sigma_i )$ the row index and $(\sigma_{i+1}, e_{i+2} ) $ the column index. This wave function can be diagonalized using two unitary matrices, $U$ and $V$, through a singular value decomposition
 \begin{equation}
   \psi_0 (s_{i-1}\sigma_i, \sigma_{i+1}e_{i+2}) = \sum_l U_{s_{i-1}\sigma_i,l} C_{l} V_{l,\sigma_{i+1}e_{i+2} } ,
 \end{equation}
 where $C$ is the diagonal singular matrix, which is also the square root of the eigenvalue matrix of the reduced density matrix of the system or environment block. Both $U$ and $V$ are basis transformation matrices. In particular, $U$ is also the matrix that diagonalizes the reduced density matrix of the system block that is defined by tracing out all basis states in the environment block
 \begin{equation}
  \rho^{\mathrm{sys}} = \mathrm{Tr}_{\mathrm{env}} | \psi_0 \rangle \langle \psi_0 | .
 \end{equation}
 Similarly, $V$ is the matrix that diagonalizes the reduced density matrix of the environment block.

 In the MPS language, $U$ and $V$ are represented by two three-leg tensors
 \begin{eqnarray}
 A^i_{s_{i-1} s_i} [\sigma_i] &=& U_{s_{i-1}\sigma_i,s_i},
 \label{Eq:Left}
 \\
 B^{i+1}_{ e_{i+1} e_{i+2} } [ \sigma_{i+1}] &=&
 V_{e_{i+1} ,\sigma_{i+1} e_{i+2} }.
 \label{Eq:Right}
 \end{eqnarray}
 After truncating the basis space to retain the largest $D$ singular values, $A^i$ and $B^{i+1}$ become left and right isometric, respectively. The ground state then becomes
 \begin{equation}
   |\psi_0  \rangle \approx \sum_{s_i e_{i+1}} C_{s_i} \delta_{s_i,e_{i+1}} |s_i,e_{i+1}\rangle ,
   \label{Eq:Ground}
 \end{equation}
 where $s_i$ and $e_{i+1}$ are the basis states retained after truncation
 \begin{eqnarray}
  |s_i \rangle &=& \sum_{s_{i-1}\sigma_i} A^i_{s_{i-1} s_i} [\sigma_i] |s_{i-1} \sigma_i\rangle ,
  \label{Eq:sys}
  \\
  |e_{i+1} \rangle &=& \sum_{e_{i+2}\sigma_{i+1}} B^{i+1}_{ e_{i+1} e_{i+2} } [ \sigma_{i+1}] |\sigma_{i+1} e_{i+2} \rangle .
  \label{Eq:env}
 \end{eqnarray}

 Equations (\ref{Eq:sys}) and (\ref{Eq:env}) hold recursively for all the lattices in the system and environment blocks, respectively. Substituting them into Eq. (\ref{Eq:Ground}) recursively, we can eventually express $\psi_0$ as an MPS
 \begin{equation}
 \psi_0 = A^1 [ \sigma_{1} ]\ldots A^{i}[\sigma_{i}] C B^{i+1}[ \sigma_{i+1} ] \ldots B^L[ \sigma_L ] ,
 \label{Eq:MPS_bond}
 \end{equation}
 Graphically, it can be represented as
 \begin{eqnarray}
  \psi_0 =
  \begin{array}{l}
  \begin{tikzpicture}[every node/.style={scale=1},scale=.35]
    \LeftIsometry{-6.4}{-0.5}{1}{1}{\sigma_1}{A_1}
    \draw (-3.75,0) node {$\cdots$} ;
    \LeftIsometry{-2.4}{-0.5}{1}{1}{\sigma_i}{A_i}
    \draw (0,0) circle (0.5) node {$C$} ;
    \RightIsometry{2.4}{-0.5}{1}{1}{\sigma_{i+1}}{B_{i+1}}
    \draw (3.85,0) node {$\cdots$} ;
    \RightIsometry{6.4}{-0.5}{1}{1}{\sigma_L}{B_L}
  \end{tikzpicture}
  \end{array}
  .
 \end{eqnarray}
 To improve the accuracy of the ground state wave function, one can increase the bond dimension $D$. Alternatively, one can also use several different MPS, not necessarily orthogonal to each other, to represent the ground state wave function \cite{Huang_2018}.

 If  $M>1$ eigenstates are targeted, we obtain $M$ orthonormalized eigenfunctions, $\psi_m (s_{i-1}, \sigma_i ,\sigma_{i+1}, e_{i+2})$ ($m=0,\ldots , M-1$), by diagonalizing the renormalized Hamiltonian at each step of DMRG iteration. Again, we can find a unitary matrix $U$ to diagonalize the reduced density matrix of the system. But the reduced density matrix is now defined by
 \begin{equation}
  \rho^{\mathrm{sys}} = \sum_m w_m \mathrm{Tr}_{\mathrm{env}} | \psi_m \rangle \langle \psi_m | ,
 \end{equation}
 where $w_m$ is a positive weighting factor whose sum equals 1. Similarly, we can find another unitary matrix $V$ to diagonalize the reduced density matrix of the environment block.

 In this case, the $m$th eigenstate can be represented as
  \begin{equation}
   |\psi_m  \rangle \approx \sum_{s_i e_{i+1}} C_{s_i,e_{i+1}}[m] |s_i,e_{i+1}\rangle ,
   \label{Eq:Excited}
 \end{equation}
 where $C[m]$ is a matrix defined by
  \begin{eqnarray}
   C_{s_i,e_{i+1}}[m] &=& \sum_{s_{i-1}\sigma_i }  \sum_{ \sigma_{i+1} e_{i+2}} \psi_m (s_{i-1} \sigma_i \sigma_{i+1} e_{i+2}) \nonumber \\
   && U^*(s_{i-1} \sigma_i,s_i) V^*(e_{i+1}, \sigma_{i+1} e_{i+2}) .
 \end{eqnarray}
 
 Following the steps leading to the MPS representation of the ground state and using the recursive relations (\ref{Eq:sys}) and (\ref{Eq:env}), we can also express $\psi_m$ as an MPS:~\cite{Xiang2023}
 \begin{equation}
 \psi_m = A^1 [ \sigma_{1} ]\ldots A^{i}[\sigma_{i}] C[m] B^{i+1}[ \sigma_{i+1} ] \ldots B^L[ \sigma_L ] .
 \end{equation}
 The corresponding graphical representation is
\begin{eqnarray}
  \psi_{m} =
  \begin{array}{l}
  \begin{tikzpicture}[every node/.style={scale=1},scale=.4]
    \LeftIsometry{-6.4}{-0.5}{1}{1}{\sigma_1}{A^1}
    \draw (-3.75,0) node {$\cdots$} ;
    \LeftIsometry{-2.4}{-0.5}{1}{1}{\sigma_{i}}{A^{i}}
    \draw (0,0) circle (0.5) node {$C$} ;
    \RightIsometry{2.4}{-0.5}{1}{1}{\sigma_{i+1}}{B^{i+1}}
    \draw (3.85,0) node {$\cdots$} ;
    \RightIsometry{6.4}{-0.5}{1}{1}{\sigma_L}{B^L}
%    \draw (0,0.5) -- (0,1.25) node[above] {$\sigma_i$} ;
    \draw[snake=coil,segment length=2pt,segment amplitude=1pt] (0,-0.5) -- (0,-1.25) node[below] {$m$} ;
  \end{tikzpicture}
  \end{array}
  . \label{Eq:mps-C}
\end{eqnarray}
 Here we use the same symbol $C$ to represent the central canonical tensor, but $C$ now contains an extra leg ``$m$''.

 As $A^j$ and $B^j$ are left and right canonicalized,
 \begin{equation}
    \sum_{\sigma} A^{j\dagger} [\sigma] A^j [\sigma] =
    \mathbbm{1}, \quad
     \sum_{\sigma} B^j[\sigma] B^{j\dagger}[\sigma ] =
    \mathbbm{1} ,  \label{Eq:Canonical}
\end{equation}
 it is simple to show that $\psi_m$ are orthonormalize
 \begin{equation}
   \langle \psi_{m^\prime} |\psi_m \rangle  = \delta_{m,m^\prime} ,
   \label{Eq:orthonormal}
 \end{equation}
 if $C[m]$ is orthonormalized:
 \begin{equation}
 \begin{array}{l}
 \begin{tikzpicture}[every node/.style={scale=1},scale=0.5]
   \draw (0,-0.8) circle (0.5) node {$C$}
        (0,0.8) circle (0.5) node {$C^*$} ;
   \draw[rounded corners] (-0.5,-0.8) -- (-1.5,-0.8) -- (-1.5,0.8) -- (-0.5,0.8)
                         (0.5,-0.8) -- (1.5,-0.8) -- (1.5,0.8) -- (0.5,0.8) ;
   \draw[snake=coil,segment length=2pt,segment amplitude=1pt] (0,-1.3) -- (0,-2) node[left] {$m$} (0,1.3) -- (0,2) node[left] {$m^\prime$} ;
 \end{tikzpicture}
 \end{array}
    = \delta_{m,m^{\prime}} .
    \label{Eq:BondCenter}
\end{equation}

 In the above expression, the canonical center $C[m]$ is defined on the bond linking sites $i$ and $i+1$. One can also absorb $C[m]$ into $A^i$ and define a canonical center at site $i$
 \begin{equation}
   \begin{array}{l}
   \begin{tikzpicture}[every node/.style={scale=1},scale=.5]
     \draw (-6.25,0) circle (0.5) node {$C^i$}
           (-6.25,0.5) -- (-6.25,1.25)  % node[above] {$\sigma_i$}
           (-6.75,0) -- (-7.5,0)
           (-5.75,0) -- (-5,0)
           (-4,-0.05) node {$=$} ;
    \draw[snake=coil,segment length=2pt,segment amplitude=1pt] (-6.25,-0.5) -- (-6.25,-1.25)  node[left] {$m$} ;

     \LeftIsometry{-2.4}{-0.5}{1}{1}{}{}
     \draw (-1.9,0) node {$A^i$} ;

     \draw (0,0) circle (0.5) node {$C$}
           (-0.5,0) -- (-1.25,0)
           (0.5,0) -- (1.25,0) ;
    \draw[snake=coil,segment length=2pt,segment amplitude=1pt] (0,-0.5) -- (0,-1.25) node[left] {$m$} ;
   \end{tikzpicture}
   \end{array}.
 \end{equation}

 In this case, $\psi_m$ becomes
\begin{equation}
  \psi_{m} =
  \begin{array}{l}
  \begin{tikzpicture}[every node/.style={scale=1},scale=.43]
    \LeftIsometry{-6.4}{-0.5}{1}{1}{\sigma_1}{A^1}
    \draw (-3.75,0) node {$\cdots$} ;
    \LeftIsometry{-2.4}{-0.5}{1}{1}{\sigma_{i-1}}{A^{i-1}}
    \draw (0,0) circle (0.5) node {$C^i$} ;
    \RightIsometry{2.4}{-0.5}{1}{1}{\sigma_{i+1}}{B^{i+1}}
    \draw (3.85,0) node {$\cdots$} ;
    \RightIsometry{6.4}{-0.5}{1}{1}{\sigma_L}{B^L}
    \draw (0,0.5) -- (0,1.25) node[above] {$\sigma_i$} ;
    \draw[snake=coil,segment length=2pt,segment amplitude=1pt] (0,-0.5) -- (0,-1.25) node[below] {$m$} ;
  \end{tikzpicture}
  \end{array}
  . \label{Eq:SiteCenter}
 \end{equation}
 This is just the bundled MPS introduced in Ref. [\onlinecite{senechal2022direct}]. The canonical center $C^i$ is now a four-leg tensor and the orthonormal condition of $\psi_m$ becomes
\begin{equation}
  \begin{array}{l}
  \begin{tikzpicture}[every node/.style={scale=1},scale=0.5]
   \draw (0,-0.8) circle (0.5) node {$C_i$}
         (0,0.8) circle (0.5) node {$C_i^*$}
         (0,-0.3) -- (0, 0.3) ;
   \draw[rounded corners] (-0.5,-0.8) -- (-1.5,-0.8) -- (-1.5,0.8) -- (-0.5,0.8)
                         (0.5,-0.8) -- (1.5,-0.8) -- (1.5,0.8) -- (0.5,0.8) ;
   \draw[snake=coil,segment length=2pt,segment amplitude=1pt] (0,-1.3) -- (0,-2) node[left] {$m$} (0,1.3) -- (0,2) node[left] {$m^\prime$} ;
 \end{tikzpicture}
 \end{array}
    = \delta_{m,m^{\prime}}.
    \label{Eq:mmps-2}
\end{equation}

\section{Methods}

\subsection{Multi-target update (MTU)}

 The MTU is essentially a canonicalization method that updates relevant local tensors without explicitly contracting the whole MPS. Like TEBD, this method is particularly suitable for studying a system with short-range interactions. As an example, let us consider a Hamiltonian with nearest-neighbor interactions $H_{i, i+1}$ only:
 \begin{equation}
   H = \sum_{i=1}^{L-1} H_{i, i+1} .
 \end{equation}
 To find the low-energy Hilbert space that optimizes the multi-target MPS, we iteratively apply the projection operator $\exp (-\tau H)$ to these MPS. Here, $\tau$ is a small parameter that is used to decouple $\exp (-\tau H)$ into a product of local projection operators, $\exp (-\tau H_{i,i+1})$, through the second-order Trotter-Suzuki decomposition formula
\begin{equation}
    e^{-\tau H} = \prod_i e^{-\tau H_{i,i+1}} + O(\tau^2) .
\end{equation}
 In practical calculation, we sweep the lattice by applying the local projection operators to MPS alternatively from one end to the other end. At each step, the canonical center is moved one site along the direction of the sweep.

 We use the MPS states $\psi_m$ represented by Eq.~(\ref{Eq:SiteCenter}) to demonstrate how to update the canonical center and other local tensors when we sweep the lattice from left to right. By applying the local projection operator $\exp (-\tau H_{i,i+1})$ to Eq.~(\ref{Eq:SiteCenter}), this changes  $\psi_m$ to
\begin{equation}
  \psi^\prime_{m} =
  \begin{array}{l}
  \begin{tikzpicture}[every node/.style={scale=1},scale=.4]
    \LeftIsometry{-7}{-0.5}{1}{1}{\sigma_1}{A^1}
    \draw (-4.35,0) node {$\cdots$} ;
    \LeftIsometry{-3}{-0.5}{1}{1}{\sigma_{i-1}}{A^{i-1}}
    \draw (-1.1,-0.5) rectangle (1.1,0.5)
          (0,0) node {$T$}
          (-0.75,0.5) -- (-0.75,1.25) node[above] {$\sigma_i$}
          (0.75,0.5) -- (0.75,1.25) node[above] {$\sigma_{i+1}$};
    \RightIsometry{3.}{-0.5}{1}{1}{\sigma_{i+2}}{B^{i+2}}
    \draw (4.45,0) node {$\cdots$} ;
    \RightIsometry{7}{-0.5}{1}{1}{\sigma_L}{B^L}
    \draw[snake=coil,segment length=2pt,segment amplitude=1pt] (0,-0.5) -- (0,-1.25) node[below] {$m$} ;
  \end{tikzpicture}
  \end{array}
  ,
\end{equation}
where $T$ is a five-leg tensor:
\begin{equation}
  \begin{array}{l}
  \begin{tikzpicture}[every node/.style={scale=1},scale=.4]
     \draw (-1.1,-0.5) rectangle (1.1,0.5)
          (0,0) node {$T$}
          (-0.75,0.5) -- (-0.75,1.25) % node[above] {$\sigma_i$}
          (0.75,0.5) -- (0.75,1.25) % node[above] {$\sigma_{i+1}$}
          (-1.1,0) -- (-1.85,0) % node[left] {$\alpha$}
          (1.1,0) -- (1.85,0) ; % node[right] {$\beta$} ;
    \draw[snake=coil,segment length=2pt,segment amplitude=1pt] (0,-0.5) -- (0,-1.25) ; % node[below] {$m$} ;
    \draw (0,2.15) node {} (3,-0.05) node {$=$} ;
  \end{tikzpicture}
  \end{array}
\begin{array}{l}
\begin{tikzpicture}[every node/.style={scale=1},scale=0.5]
%  \draw (-2,-0.05) node[left] {$T^m_{\alpha\sigma_i , \beta\sigma_{i+1}} = $} ;
  \RightIsometry{2.4}{-0.5}{1}{1}{}{B^{i+1}}
  \fill[white] (1.8,1) rectangle (2,1.25) ;
  \draw (0,0) circle (0.5) node {$C^i$} ;
  \draw (-0.5,0) -- (-1.25,0) %node[left] {$\alpha$}
        (0,0.5) -- (0,1.)
        % (3.15,0) node[right] {$\beta$}
        (0,2) -- (0,2.5) % node[above] {$\sigma_i$}
        (1.9,2)--(1.9,2.5) ;% node[above] {$\sigma_{i+1}$} ;
  \draw[snake=coil,segment length=2pt,segment amplitude=1pt] (0,-0.5) -- (0,-1.25) ; % node[below] {$m$} ;
  \draw[rounded corners] (-1,1) -- (-1,2) -- (3,2) -- (3,1) -- cycle ;
  \draw (1,1.5) node {$e^{-\tau H_{i,i+1}}$} ;
\end{tikzpicture}
\end{array}
\end{equation}

Taking a QR decomposition to decouple $T$ into a product of a unitary matrix $A_i$ and an upper triangular matrix $\tilde C_{i+1}$, we obtain
\begin{equation}
  \begin{array}{l}
  \begin{tikzpicture}[every node/.style={scale=1},scale=.35]
     \draw (-1.1,-0.5) rectangle (1.1,0.5)
          (0,0) node {$T$}
          (-0.75,0.5) -- (-0.75,1.25)  node[above] {$\sigma_i$}
          (0.75,0.5) -- (0.75,1.25)  node[above] {$\sigma_{i+1}$}
          (-1.1,0) -- (-1.85,0) % node[left] {$\alpha$}
          (1.1,0) -- (1.85,0) ; % node[right] {$\beta$} ;
    \draw[snake=coil,segment length=2pt,segment amplitude=1pt] (0,-0.5) -- (0,-1.25) node[below] {$m$} ;
    \draw (2.9,-0.05) node {$=$} ;
  \end{tikzpicture}
  \end{array}
  \begin{array}{l}
  \begin{tikzpicture}[every node/.style={scale=1},scale=.35]
     \draw[rounded corners] (-0.5,-0.5) rectangle (0.5,0.5) ;
     \draw (-0.5,0) -- (-1.25,0)
           (0,0.5) -- (0,1.25) node[above] {$\sigma_i$}
           (0,-0.5) node [below] {$A^i$} ;
     \draw[very thick] (0.5,0) -- (1.25,0) ;
     \draw (1.75,0) circle (0.5)
           (3.,-0.15) node[below] {$\tilde C^{i+1}$}
           (2.25,0) -- (3.25,0)
           (1.75,0.5) -- (1.75,1.25) node[above] {$\sigma_{i+1}$} ;
    \draw[snake=coil,segment length=2pt,segment amplitude=1pt] (1.75,-0.5) -- (1.75,-1.25) node[below] {$m$} ;
    \draw (4.4,-0.05) node {$\approx$} ;
  \end{tikzpicture}
  \end{array}
    \begin{array}{l}
  \begin{tikzpicture}[every node/.style={scale=1},scale=.35]
     \LeftIsometry{-0.45}{-0.5}{1}{1}{\sigma_i}{A^{i}}
     \draw (0.7,0) -- (1.5,0) ;
     \draw (2,0) circle (0.5)
           (3.25,-0.15) node[below] {$\tilde C^{i+1}$}
           (2.5,0) -- (3.25,0)
           (2,0.5) -- (2,1.25) node[above] {$\sigma_{i+1}$} ;
    \draw[snake=coil,segment length=2pt,segment amplitude=1pt] (2,-0.5) -- (2,-1.25) node[below] {$m$} ;
    \draw (4,0) node {.} ;
  \end{tikzpicture}
  \end{array}
  \label{Eq:T_decouple}
\end{equation}
  Here a thick bond is used to emphasize that it is a bond before truncation. After truncation (indicated by the approximate equality), it becomes a thin bond whose bond dimension is lower than the thick one. This truncation introduces errors in the determination of the multi-target MPS. However, these truncation errors are not accumulated in the sweep and therefore do not affect the accuracy of the final converged results.

  Substituting Eq.~(\ref{Eq:T_decouple}) into Eq.~(\ref{Eq:SiteCenter}), we obtain an updated MPS whose canonical center moves to site $i+1$. However, the canonical center $\tilde C_{i+1}$ obtained is not orthonormalized. Consequently, the updated MPS $\psi^\prime_m$ are also not orthonormalized. Before taking the step of projection, we should reorthonormalize these MPS. For doing this, we first diagonalize the following matrix
\begin{equation}
 X_{m^\prime , m} =
\begin{array}{l}
\begin{tikzpicture}[every node/.style={scale=1},scale=0.5]
  \draw (0,-0.8) circle (0.5)
        (1.1,-0.9) node[below] {$\tilde C^{i+1}$}
        (0,0.8) circle (0.5)
        (1.2,1) node[above] {$\tilde C^{i+1 *}$}
        (0,-0.3) -- (0, 0.3) ;
  \draw[rounded corners] (-0.5,-0.8) -- (-1.5,-0.8) -- (-1.5,0.8) -- (-0.5,0.8)
                         (0.5,-0.8) -- (1.5,-0.8) -- (1.5,0.8) -- (0.5,0.8) ;
  \draw[snake=coil,segment length=2pt,segment amplitude=1pt] (0,-1.3) -- (0,-2) node[left] {$m$} (0,1.3) -- (0,2) node[left] {$m^\prime$} ;
\end{tikzpicture}
\end{array}
   =\sum_n U_{n,m^\prime}^* \lambda_n U_{n,m},
   \label{Eq:Orth_1}
\end{equation}
 where $U$ is a unitary matrix and $\lambda_n$ is the eigenvalue of $X$. We then update the canonical tensor at site $i+1$ by the formula:
\begin{equation}
  C^{i+1}_{\alpha \beta}[\sigma, m]  = \sum_n \tilde C^{i+1}_{\alpha \beta}[\sigma, n]  U^*_{m,n} \lambda_m^{-1/2} . \label{Eq:Orth_2}
\end{equation}
 It is straightforward to show that $C^{i+1}$ such defined is orthonormalized.

 The above canonicalization steps (\ref{Eq:Orth_1}) and (\ref{Eq:Orth_2}) are crucial to maintaining the orthonormal properties of the multi-target MPS. Repeating the above steps by sweeping the lattice for sufficiently many times, we will cool down the temperature and project $\psi_m$ onto the subspace spanned by the $M$-lowest eigenstates of $H$ approximately.

 The converged MPS $\psi_m$ are not automatically the eigenstates of the Hamiltonian. To find the eigenfunctions, we first calculate the matrix elements of $H$ in the subspace spanned by these $M$ orthonormal basis states $\psi_m$:
 \begin{equation}
  H_{m,n} = \langle \psi_m | H | \psi_n\rangle.
  \label{Eq:Ham}
 \end{equation}
 The eigenvalues and eigenvectors of this Hamiltonian give the approximate solution of the lowest-$M$ eigenenergies and the corresponding eigenstates of the system. 

 To carry out the above projection efficiently, it is suggested not to start with a too small $\tau$. Instead, one should gradually decrease $\tau$ to reduce the Trotter error after completing several sweeps.

\subsection{Variational Riemannian optimization (VRO)}

 One can also take the tensor elements of the multi-target MPS as variational parameters to determine them by optimizing a cost function that implements the variational principle~\cite{PhysRevLett.93.227205, PhysRevX.9.031041}. This can improve the accuracy of results without introducing the Trotter error. A generalized variational principle~\cite{PhysRevA.37.2805} states that the sum of the energy expectation values of $M$ orthonormal basis states $\psi_m$ ($m=0,\ldots , M-1$) is always higher than or equal to the sum of the $M$-lowest eigenenergies of the full Hamiltonian:
\begin{equation}
    \sum_{m=0}^{M-1}{\langle\psi_{m}|H|\psi_{m}\rangle} \geq \sum_{m=0}^{M-1} E_{m}^{\rm{ex}},
\end{equation}
 where $E_{m}^{\rm{ex}}$ is the exact result of the $m$th lowest eigenenergy of $H$. Thus we can define the cost function as
  \begin{equation}
    f(A,B,C) = \sum_{m=0}^{M-1}{\langle\psi_{m}|H|\psi_{m}\rangle} ,
    \label{Eq:costfunc}
  \end{equation}
  where $(A,B,C)$ stands for the left, right, and canonical center tensors in Eq.~(\ref{Eq:mps-C}) or (\ref{Eq:SiteCenter}).

  Equation (\ref{Eq:costfunc}) holds when the orthonormal condition (\ref{Eq:orthonormal}) or (\ref{Eq:mmps-2}) is valid. To determine the values of variational parameters, we should therefore maintain the orthonormality of $\psi_m$ in searching for the optimal path that minimizes the cost function.

  In the MPS representation of $\psi_m$, Eq.~(\ref{Eq:mps-C}), all local tensors, including $A^i$, $B^j$, and $C$, are either left or right canonicalized. With proper regrouping of tensor indices, they can all be represented as column isometric or unitary matrices. For example, we can convert the left canonical tensor $A^i_{s,s^\prime}[\sigma ]$ into an isometry by setting $(s, \sigma)$ as the row index and $s^\prime$ as the column index of an isometric matrix $W$ whose matrix elements are defined by
  \begin{equation}
    W_{s\sigma, s^\prime} = A^i_{s,s^\prime}[\sigma] .
  \end{equation}
  The row dimension of $W$ is not less than its column dimension. Assuming it to be a $n \times p$ matrix with $n\ge p$, $W$ should satisfy the constraint
  \begin{equation}
        W^\dag W = \mathbbm 1 .
    \label{Eq:Isometric}
  \end{equation}
  When $n=p$, $W$ is a unitary matrix. For convenience, we call $W$ isometric no matter whether $n>p$ or $n=p$. Similarly, one can convert the right canonical tensor $B^j_{e^\prime, e}[\sigma]$ into an isomatric matrix by taking $(e, \sigma)$ as the row index and $e^\prime$ the column index. The canonical center $C_{s,e}[m]$ is converted into an isometric matrix by taking $m$ as the column index and $(s,e)$ as the row index.

  A matrix that satisfies the constraint (\ref{Eq:Isometric}) forms a manifold, called Stiefel manifold, which is denoted as $St(n,p)$. This kind of matrices widely appears in singular value decompositions \cite{sato2013riemannian}, image processing \cite{wei2022neighborhood, cui2022taotf}, the linear eigenvalue problem \cite{Wen2016,saad2011numerical}, the Kohn-Sham total energy minimization \cite{zhang2014gradient,altmann2022energy}, and tensor-network representations of quantum states~\cite{PhysRevLett.99.220405, Luchnikov_2021, PhysRevLett.124.037201, PhysRevResearch.3.023236}.

  A matrix $W$ in the Stiefel manifold remains in that manifold if it is right multiplied by a $p\times p$ unitary matrix $U$:
  \begin{eqnarray}
    \tilde W = WU.  \label{Eq:redefine}
  \end{eqnarray}
  In the isometric tensor network, such a unitary matrix $U$ can be interpreted as a gauge transformation on the bond corresponding to the column of $W$. It implies that there is a gauge redundancy in determining an isometric matrix $W$ since another unitary matrix $U^\dagger$, whose product with $U$ forms an identity, can be absorbed into the tensor on the other end of the bond. We call two matrices, $W$ and $\tilde W$, equivalent if they are related to each other by a unitary transformation defined by Eq.~(\ref{Eq:redefine}).

  To remove the gauge ambiguity, we introduce the Grassmann manifold which is defined as the quotient manifold of the Stiefel manifold under the equivalence relation and represented as $Gr(n, p) = St(n, p) / U(p)$, where $U(p)$ is a unitary manifold of dimension $p$. The dimension of the Grassmann manifold is $(n-p)p$.

  Clearly, minimizing the cost function $f(A,B,C)$ can be reformulated as an optimization problem that minimizes local isometric matrices $W$ that satisfy the constraint (\ref{Eq:Isometric}). This constrained optimization is a highly  nonlinear problem. A promising approach to solving this nonlinear problem is to only target one local tensor while keeping all other tensors fixed in the minimization of the cost function and to sweep over all local tensors iteratively. At each step, on the other hand, the local tensor is determined by the Riemannian optimization \cite{absil2008optimization}. This approach optimizes a local tensor on the Riemannian manifold, including the Stiefel manifold as well as the Grassmann manifold, by retracting the travel vector in tangent space generated from the cost function to a point on the manifold~\cite{Zhu2017, zhang2016riemannian, hu2018adaptive}.

  The Riemannian optimization starts with a vector  $X$ in the tangent space of the Grassmann manifold. To ensure the resulting matrix after a move along that direction, i.e.,  $W + \varepsilon X$ with $\varepsilon$ a moving step parameter, to remain isometric to the first order in $\varepsilon$, it is simple to show that $X$ should satisfy the equation 
  \begin{eqnarray} \label{Tangent Vector}
       X = WQ + W_\perp R,
  \end{eqnarray}
  where $Q$ is an antihermitian matrix, $Q = -Q^\dag$, $W_\perp$ is a unitary complement of $W$, satisfying the equation
  \begin{equation}
    WW^\dag + W_\perp W_\perp^\dag = \mathbbm 1 ,
  \end{equation}
  and $R$ is an arbitrary $(n - p) \times p$ matrix. Furthermore, if $W$ is a point in the Grassmann manifold, a unitary gauge transformation can be imposed to ensure $Q = 0$. This yields
  \begin{equation}
    X = W_\perp R. \label{Eq:Grassmann}
  \end{equation}

  To determine the optimal vector on the tangent space of the Grassmann manifold, we first calculate the derivative of the cost function without considering the constraint (\ref{Eq:Isometric}), $D = \partial_W f(W)$. However, this derivative contains both the components on and those not on the tangent space of the Grassmann manifold of $W$. Using Eq.~(\ref{Eq:Grassmann}) and the properties of the tangent vectors, it can be shown that the components of $D$ on the tangent space of the Grassmann manifold are given by
  \begin{equation}
    G = \left( 1 - WW^\dag \right) D .  \label{PartDeriv}
  \end{equation}
  This vector represents the direction of a local optimization path, and here we adopt the Euclidean metric
  \begin{equation}
    R(X,Y) = \mathrm{Re} \,\mathrm{Tr} (X^\dagger Y) .
  \end{equation}
  
  The Riemannian optimization finds the optimal $W$ by generating a sequential path using the tangent vectors obtained with Eq.~(\ref{PartDeriv}). Let us assume $W_k$ and $g_k$ to be a point in the Grassmann manifold and a vector in tangent space along which $W_k$ is updated at the $k$th step according to the formula
  \begin{eqnarray}
     W_{k + 1} = W_k + \varepsilon_k g_k,
     \label{update}
  \end{eqnarray}
  where $\varepsilon_k$ is the $k$th step parameter. In the steepest descent method, $g_k$ just equals the minus of the tangent vector $G_k$ of $W_k$ that is determined by Eq.~(\ref{PartDeriv}), hence $g_k = -G_k$. However, $W_{k + 1}$ may not fall onto the Grassmann manifold automatically. A retraction should be done to map it back to the Grassmann manifold.

  There are several approaches to retract $W_{k+1}$ back to the Grassmann manifold \cite{absil2008optimization, nishimori2005learning, Zhu2017}. The approach we adopt is
\begin{equation}
   W_{k+1} = \left\{
   \begin{array}{ll}
      e^{\varepsilon_k Q_k }W_k , & p \ge n/2 \\
      W_k + \varepsilon_k U_k \displaystyle \frac{1}{I - M_{k,2}} M_{k,1}, & p < n/2
      \end{array} ,
      \right. \label{Eq:WW2}
\end{equation}
where
  \begin{eqnarray}
    Q_k &=& g_k W_k^\dag - W_k g_k^\dag , \nonumber \\
    U_k &=& [g_k,W_k], \nonumber\\
    V_k &=& [W_k,-g_k] ,  \\
    M_{k,1} &=& V_k^\dag W_k, \nonumber\\
    M_{k,2} &=& \frac{\varepsilon_k}{2} V_k^\dag U_k, \nonumber
  \end{eqnarray}
  In calculating the second expression in Eq.~(\ref{Eq:WW2}), one can use the Sherman-Morrison Woodbury formula \cite{Zhu2017, Wen2013,zhu2021cayley} to reduce the computational complexity from $O(n^3)$ to $O(np^2 + p^3)$.

  However, the steepest descent may not be the best approach in optimization. If, instead, the conjugate gradient method is used, the searching direction of the $(k+1)$-step should depend on the searching direction of one step before, hence
  \begin{eqnarray}
    g_{k + 1} = -G_{k + 1} + \beta_{k + 1} g_k, \label{CGEulidean}
  \end{eqnarray}
  and $g_0 = -G_0$. This conjugate approach, unfortunately, does not work because the two terms on the right-hand side of Eq.~(\ref{CGEulidean}) belong to two different tangent spaces. More specifically, $g_k$ is not on the tangent space of $W_{k + 1}$. Nevertheless, this problem can be removed by introducing a vector transport $\mathcal T$ to map $g_k$ onto the tangent space of $W_{k + 1}$, which yields
  \begin{equation}
        g_{k + 1} = -G_{k + 1} + \beta_{k + 1} \mathcal T (g_k),
        \label{CGRiemannian}
  \end{equation}
  where the vector transport $\mathcal T$ is taken as the differentiation of the retraction
   \begin{equation}
    \mathcal T (g_k)=\left\{
    \begin{array}{ll}
        e^{\varepsilon_k Q_k} g_k, &  p \ge n/2 \\
        U_k \left[ M_{k,1} +\displaystyle \frac{2-M_{k,2}}{I - M_{k,2}}  M_{k,3} \right] , &  p < n/2
    \end{array}
    \right. ,
   \end{equation}
where
  \begin{equation}
        M_{k,3} =  M_{k,2} \left(I - M_{k,2}\right)^{-1} M_{k,1} .
  \end{equation}

 We determine the step parameters, $\varepsilon_k$ and $\beta_k$, using the  traditional Fletcher Reeves algorithm \cite{fletcher1964function}, but the  inner products are replaced by the metrics of the Grassmann manifold \cite{sato2021riemannian}. Particularly, $\beta_k$ is determined by the formula
  \begin{equation}
    \beta_k = \frac{R (G_k,G_k)}{R (G_{k - 1}, G_{k - 1})}.
    \label{Beta}
  \end{equation}
  There are some flexibilities in determining another parameter, $\varepsilon_k$. But it should satisfy the strong Wolfe conditions
  \begin{eqnarray}
        f(W_{k + 1}) &\leq & f(W_k) + c_1 \varepsilon_k R(G_k, g_k), \\ |R(G_{k + 1}, \mathcal T (g_k))| &\leq & c_2 |R (G_k, g_k)|,
  \end{eqnarray}
  with $0 < c_1 < c_2 < 1$.

  Once the variational parameters become converged, we can again determine the lowest $M$ eigenenergies and eigenstates by diagonalizing the matrix $H_{m,n}$ defined by Eq.~(\ref{Eq:Ham}). Clearly, the cost of this variational optimization scales linearly with the system size. If the maximum virtual bond dimension is $D$ and the physical bond dimension is $d$, then the computational cost of MTU scales as $O(M^2 D^3 d^3)$. The computational cost of VRO, on the other hand, scales as $O(D^3 d^3)$ for each left or right canonical tensor while as $O(M^2 D^2 d)$ for the canonical center.

\section{Results}

\begin{figure}
\centerline{\includegraphics[width=\linewidth]{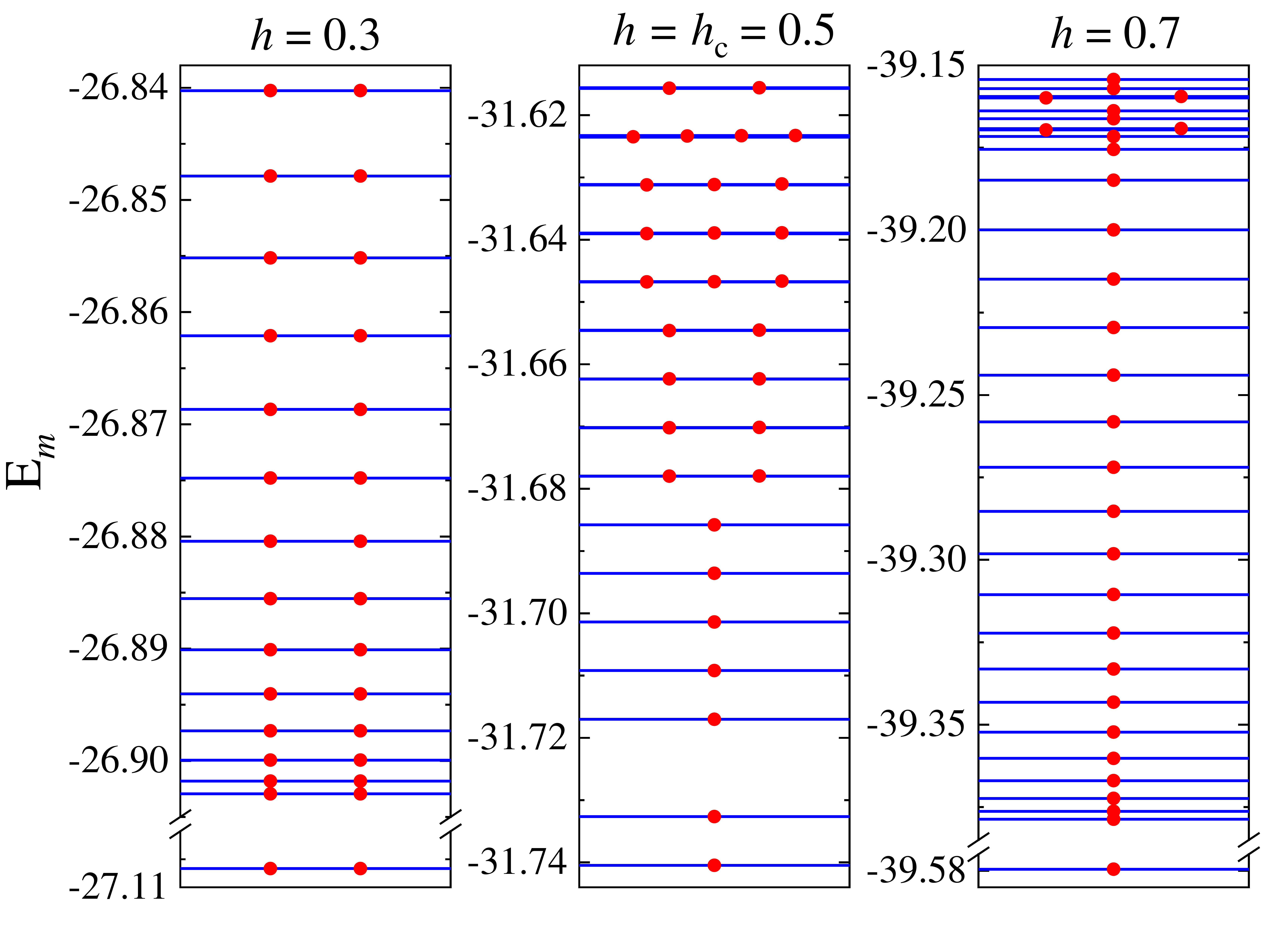}}
  \caption{Comparison of the low-energy eigenspectra (red dots) obtained by MTU with the exact solutions (blue lines) for the transverse-field Ising chain at three representative fields with $M = 30$, $L = 100$, and $D = 200$. For clarity in comparison, a break is applied to the first excited energy gap when the field is away from the critical point, and the degenerate eigenenergies obtained with MTU are separately depicted. }
	\label{Fig:Ising Spe}
\end{figure}

 We benchmark the two algorithms introduced in the preceding section using the transverse-field Ising model on a finite one-dimensional lattice with open boundary conditions:
 \begin{equation}
    H = \sum_{i=1} \left( -JS_{i}^{z}S_{i+1}^{z} + hS^{x}_{i} \right)
 \end{equation}
 where $S_{i}^{x,z}$ are the $S=1/2$ spin operators. Without loss of generality, we set $J=1$. This model undergoes a continuous transition from a paramagnetic phase to a ferromagnetic ordered phase at a critical field $h_c = 1/2$ at zero temperature.

\begin{figure}
	\centering
        \includegraphics[width=0.9\linewidth]{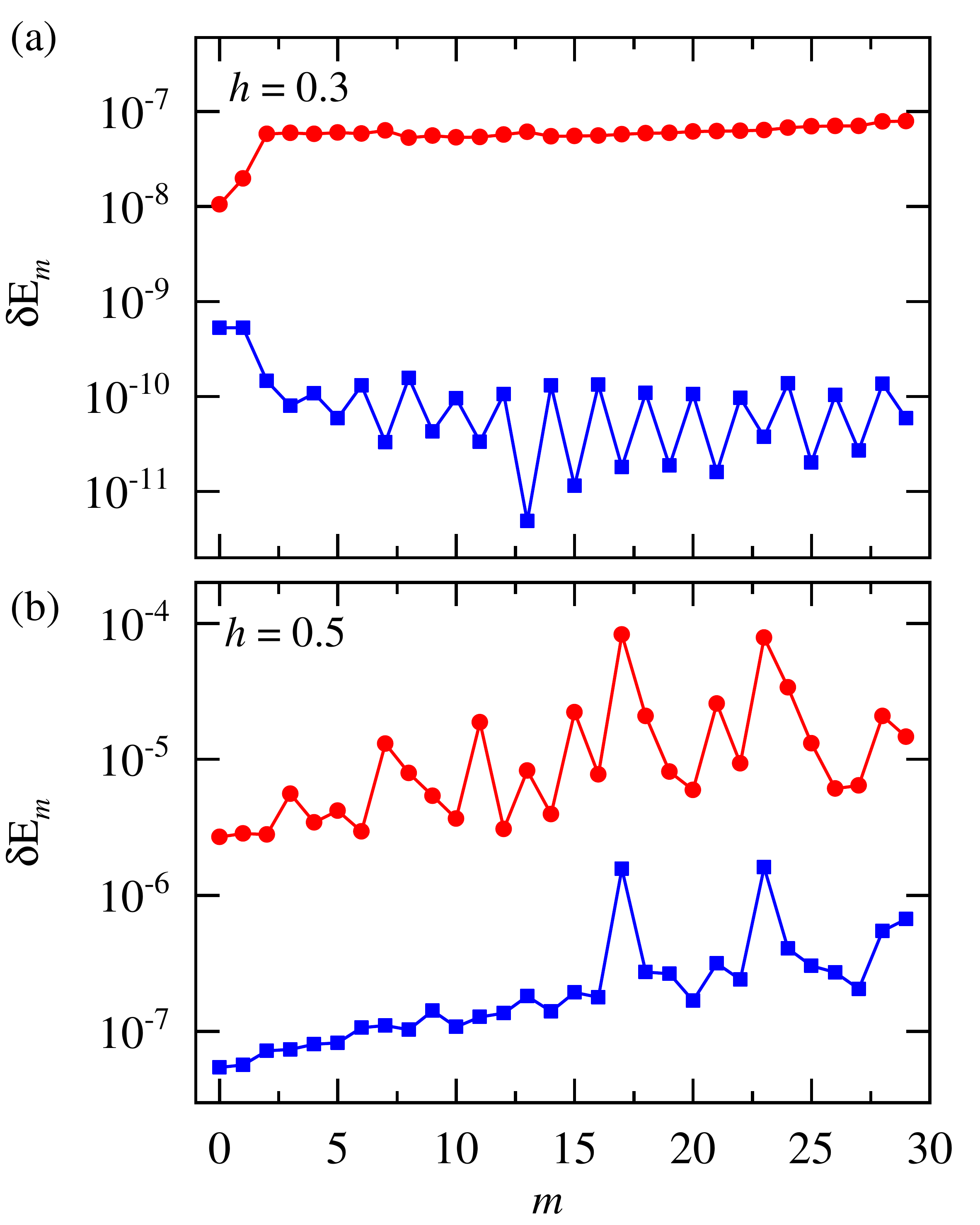}
        \caption{Absolute errors of low energy eigenspectra obtained with MTU (red dots) and VRO (blue squares) for the transverse-field Ising model at two fields with $M=30$, $L=100$ and $D=200$. The errors of the system with $h=0.7$ behave similarly to the $h = 0.3$ case.}
	\label{Fig:Ising Err}
\end{figure}

 The one-dimensional transverse-field Ising model can be converted to a noninteracting fermion model by taking the Jordan-Wigner transformation. It allows us to solve this model and calculate the full spectra exactly \cite{PFEUTY197079}. Therefore, we can make a quantitative comparison between our numerical results and the exact ones.

 Figure~\ref{Fig:Ising Spe} shows the low-energy eigenspectra obtained by MTU for the transverse-field Ising model on a lattice size $L=100$ with $M = 30$ and $D = 200$ in three representative field points. We used the same parameters as those used by Chepiga and Mila in their DMRG calculation \cite{PhysRevB.96.054425}. Our methods can accurately and efficiently calculate low-energy eigenspectra in the gapped phases as well as at the critical point. By comparison, we find that our results agree excellently with the exact ones and are systematically more accurate than those published in Ref. \cite{PhysRevB.96.054425}. Furthermore, we find that the errors in the eigenenergies in the gapped phase are smaller than that at the critical point. This differs from the observation made by Chepiga and Mila \cite{PhysRevB.96.054425}, but matches the physical expectation because a critical state bears more entanglement than a gapped state.

 To obtain the results shown in Fig.~\ref{Fig:Ising Spe}, we start with a relatively large $\tau =5$ to avoid being trapped at a local minimum of the cost function. We then gradually reduce the value of $\tau$ several times by taking roughly half of its value after several sweeps. We stop the iteration when $\tau$ reaches the order of $10^{-5} \sim 10^{-7}$ and the truncation error does not show significant change by further reducing the value of $\tau$.

\begin{figure}[htbp]
    \centering
    \includegraphics[width=0.9\linewidth]{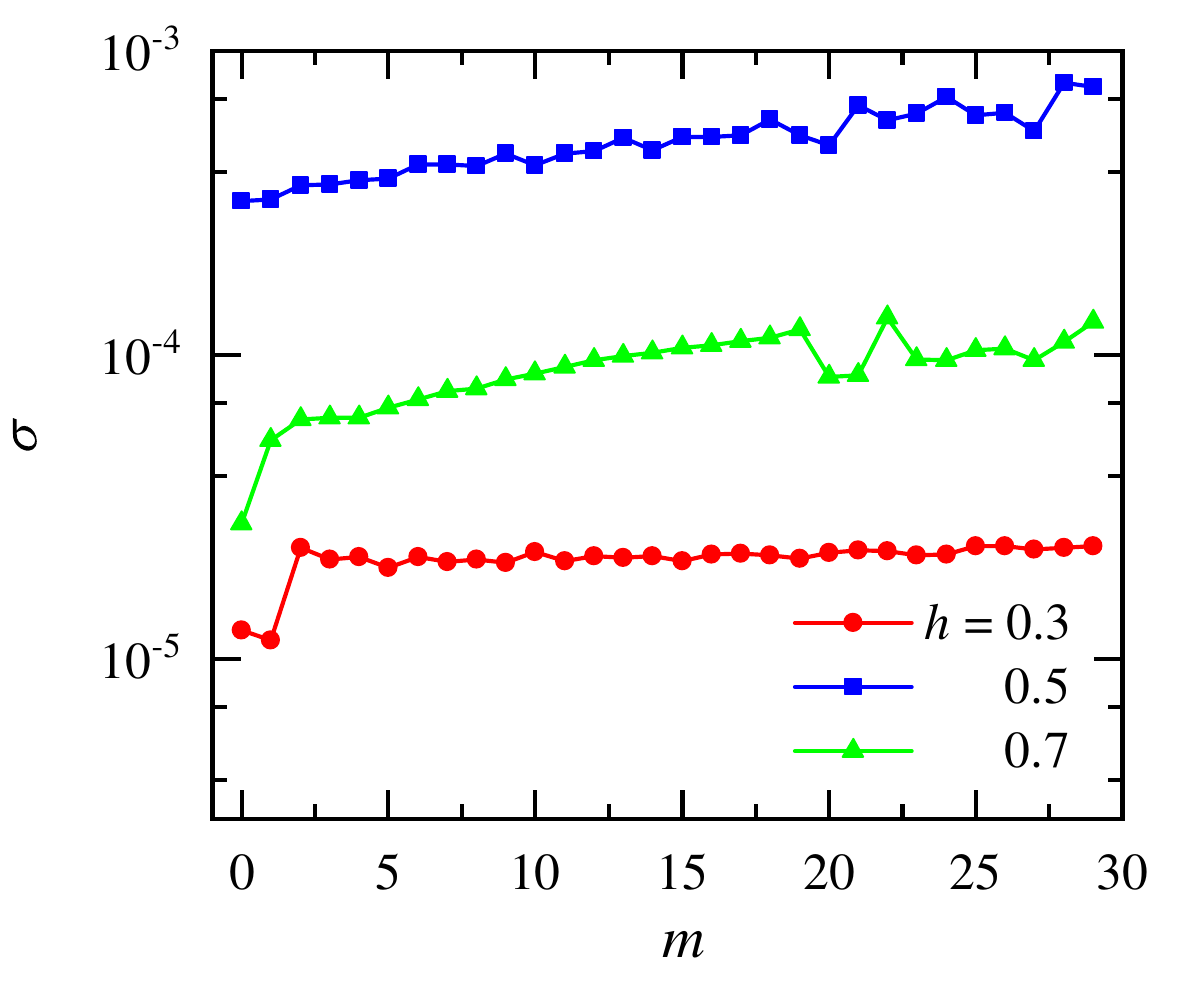}
    \caption{Variances in the eigenenergies of the transverse-field Ising model obtained by VRO with $L = 100$, $M = 30$, and $D = 200$. }
    \label{Fig:Variance}
\end{figure}

 One can use VRO to further improve the accuracy of the eigenenergies. This is because VRO does not involve a truncation step and the error comes purely from the approximation in the MPS representations of low-energy eigenstates. Furthermore, VRO can also be applied to a model with an arbitrary long-range interaction. In a VRO calculation, in principle, one can set up the initial local tensors with random numbers.  However, as the computational cost for updating the local tensors using VRO is much higher than using MTU, it is better to start a VRO calculation with an MPS first optimized by MTU whenever possible.

  Figure~\ref{Fig:Ising Err} compares the errors of the eigenenergies obtained with the two methods 
\begin{equation}
   \delta{\rm{E}}_{m} = |E_m^{\rm{ex}} - E_{m}| ,
\end{equation}
   where $E_m^{\rm{ex}}$ is the exact result of the eigenenergy.    From the calculation, we find that the absolute errors of eigenenergies obtained by VRO are $2 \sim 3$ orders of magnitude smaller than those obtained by MTU. The errors of the eigenenergies at the critical point are higher than at the field away from the critical point.

  As revealed by Fig.~\ref{Fig:Ising Err}, the errors in the eigenenergies are nearly independent of $m$ in the non-critical phases, indicating that the VRO results of the eigenenergies are uniformly converged. It further suggests that the errors in the difference between two neighboring energy levels
\begin{equation}
  \Delta_{m} = E_{m} - E_{m-1}
\end{equation}
  can be smaller than the errors in the eigenenergies in the non-critical phase. This is indeed what we find. At the critical point, however, this uniform convergence in the eigenenergies is not observed.

 \begin{figure}[htbp]
    \centering
    \includegraphics[width=0.9\linewidth]{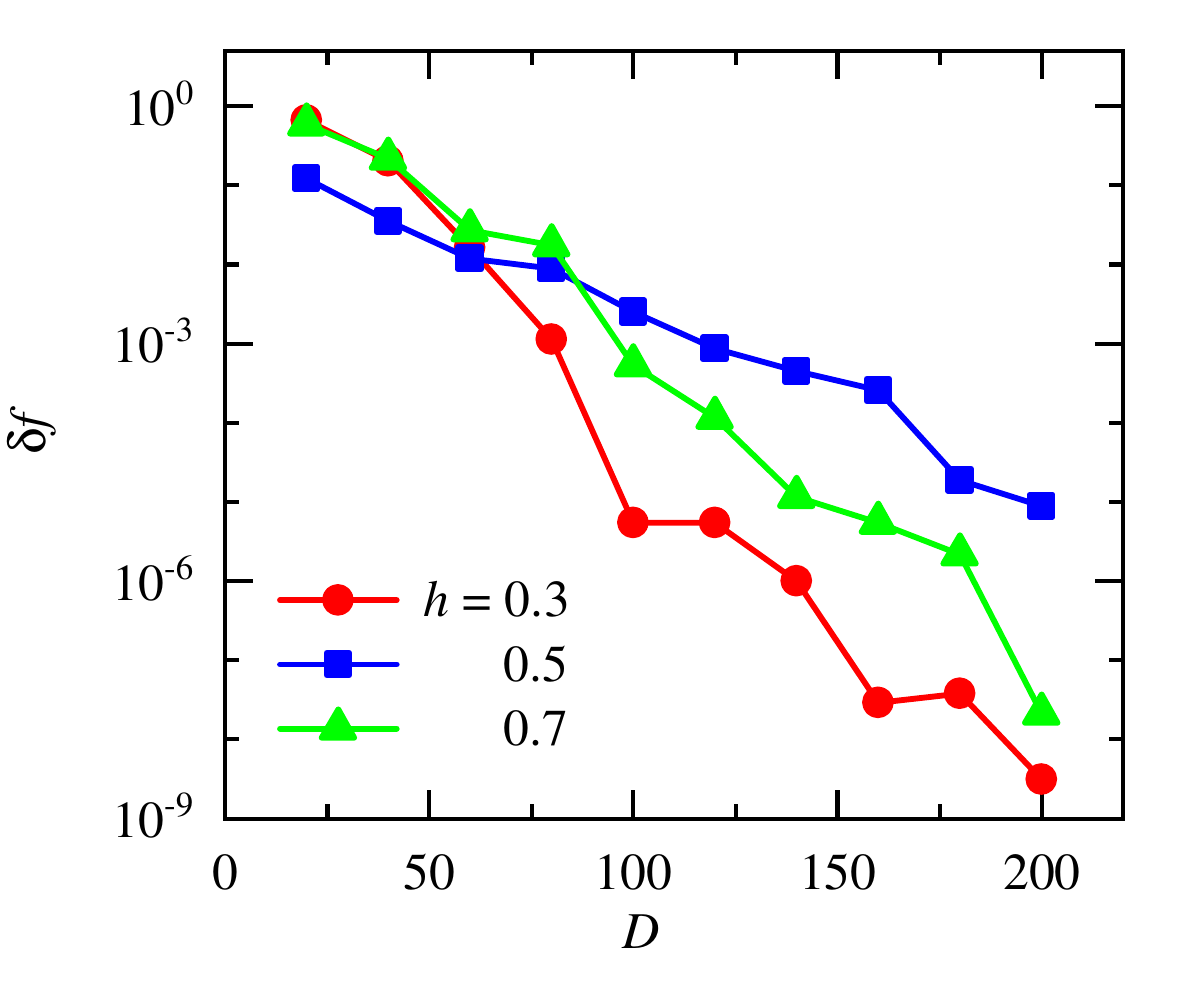}
    \caption{Absolute errors of the cost function $\delta f$ as a function of the virtual bond dimension $D$ for the transverse-field Ising model obtained by VRO. The lattice size $L=100$ and $M=30$ states are targeted. }
    \label{Fig:DiffEWithD}
\end{figure}

\begin{figure}[htbp]
    \centering
    \includegraphics[width=0.9\linewidth]{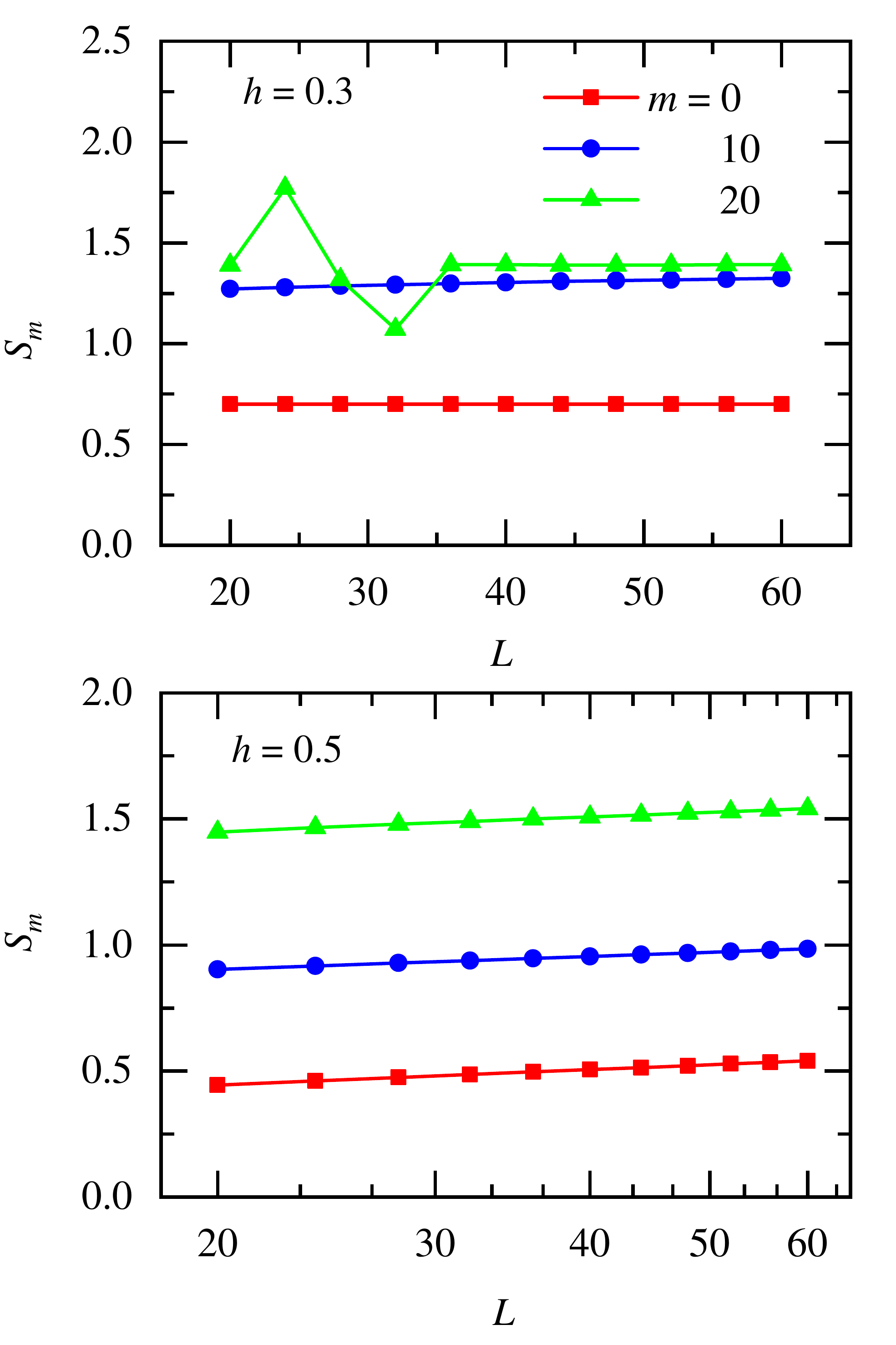}
    \caption{Entanglement entropy as a function of the lattice size $L$ for the transverse-field Ising model obtained by MTU with $h=0.3$ and $h=0.5$. The other parameters used are $D=500$ and $M=30$. }
    \label{Fig:Entropy}
\end{figure}

 \begin{figure}[htbp]
    \centering
    \includegraphics[width=0.9\linewidth]{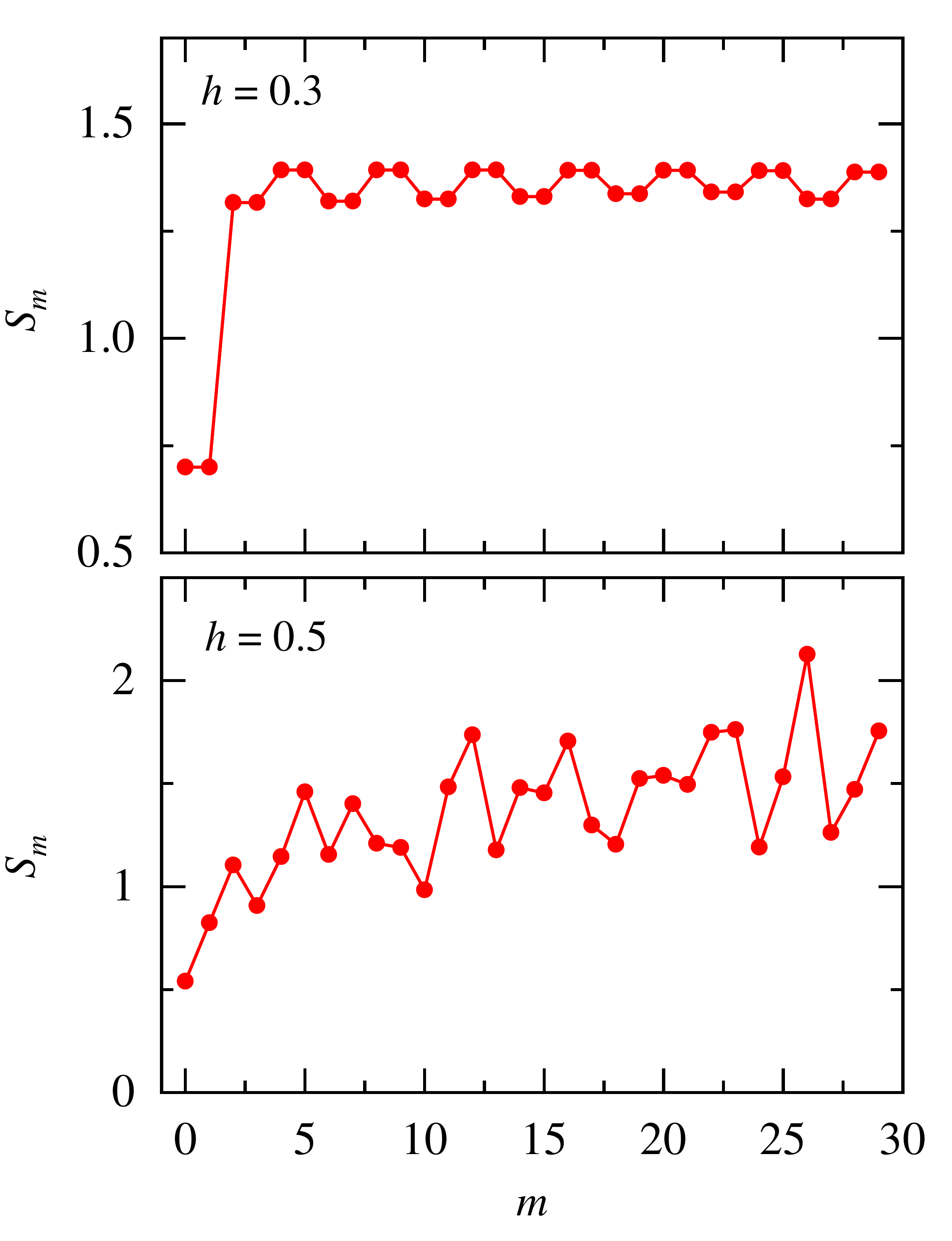}
    \caption{The entanglement entropy for the lowest $M=30$ eigenstates of the transverse-field Ising model obtained by MTU. Other parameters used are $L = 60$ and $D = 500$.}
    \label{Fig:Entropy_m}
\end{figure}

 The variances in the energy for each eigenstate
\begin{equation}
  \sigma = \sqrt{\langle H^2 \rangle - \langle H \rangle^2}
\end{equation}
 provides another measure to probe the accuracy of the results. Figure~\ref{Fig:Variance} shows the energy variance of the eigenstates calculated  by VRO for the transverse-field Ising model. As expected, the variance is smaller at the noncritical points than at the critical point. 

 Figure~\ref{Fig:DiffEWithD} illustrates how the cost function, which equals the total energy of the first $M=30$ eigenstates, converges to the exact result with the increase of the bond dimension $D$ for the transverse-field Ising model. For the three cases shown in the figure, the errors in the cost function, $\delta f = f - f^{\rm ex}$ with $f^{\rm ex}$ the exact result, drop exponentially with $D$.

 Figure~\ref{Fig:Entropy} shows how the entanglement entropy of the $m$th eigenstate
\begin{equation}
S_m = -\sum_{i=1}^D \lambda_{m,i}^2 \ln{\lambda_{m,i}^2}
\label{Entropy}
\end{equation}
 varies with the lattice size $L$ obtained by MTU. Here $\lambda_{m,i}$ is the $i$th singular value of the $m$th energy eigenfunction. At the field away from the critical point, the entanglement entropy of the low-energy eigenstates ($m = 0, 10$) converges quickly with the increase of $L$. However, for the $m=20$ eigenstate, $S_m$ oscillates severely in the small $L$ and converges when $L$ is larger than 35. In general, the entanglement entropy becomes more and more sensitive to $D$ with the increase of $m$. At critical point $h = 0.5$, $S_m$ scales logarithmically with $L$, $S_m \sim \ln L$, for all the $M=30$ eigenstates, which is consistent with the conformal field theory. 

 Figure~\ref{Fig:Entropy_m} shows the entanglement entropy as a function of $m$ for the first $M=30$ eigenstates obtained by MTU. Again the results agree accurately with the exact ones, but the errors grow with the increase of $m$. The entanglement entropies are two-fold degenerate for all the eigenstates in the case of $h = 0.3$. The entanglement entropies of the excited states alternate between two values, with a periodicity of four eigenstates. The entanglement entropy at the critical point does not show a regular pattern with particular periodicity or degeneracy.

\section{Summary}

 We show that the wave functions of the multiple target states computed by DMRG can be represented by a multi-target MPS. Leveraging this representation, we introduce two novel algorithms, MTU and VRO, for efficiently and accurately determining low-energy eigenspectra of quantum lattice models. MTU extends the commonly used TEBD method, which evaluates the ground state of a quantum system, by applying a projection operator onto the multi-target MPS to filter out the high-energy subspace. In an MTU calculation, the errors result from both the Trotter-Suzuki decomposition and the basis truncation. Reducing the value of $\tau$ can minimize the Trotter error and ensure that the truncation error, which is determined by the number of states retained, is the primary source of errors. On the other hand, VRO is a variational method that maintains the canonical form of the multi-target MPS and implements global optimization, resulting in highly accurate eigenspectra calculations. While MTU and VRO can operate independently in the study of a quantum lattice model with short-range interactions, VRO can combine with MTU to further improve the accuracy of low-energy eigenspectra.

 Using the one-dimensional transverse-field Ising model, we demonstrate the stability and accuracy of the proposed methods. In particular, our methods yield much more accurate and uniformly convergent results than DMRG \cite{PhysRevB.96.054425}, not only at the critical point but also in non-critical phases.

 % The MTU algorithm is to perform an evolution in the imaginary time direction, which projects out the low-energy subspace by cooling down the temperature. Like TEBD, MTU can also be used to perform real-time evolutions. This would allow us to calculate the time evolution of low-energy excited states. A further investigation in this direction is in progress.

\section{ACKNOWLEDGMENTS}

 This work is supported by the National Key Research and Development Project of China (Grants No.~2022YFA1403900 and No.~2017YFA0302901), the National Natural Science Foundation of China (Grants No.~11888101, No.~11874095, and No.~11974396), the Youth Innovation Promotion Association of Chinese Academy of Sciences (Grant No.~2021004), and the Strategic Priority Research Program of Chinese Academy of Sciences (Grants No.~XDB33010100 and No.~XDB33020300).

%\bibliography{ref}

\begin{thebibliography}{74}%
\makeatletter
\providecommand \@ifxundefined [1]{%
 \@ifx{#1\undefined}
}%
\providecommand \@ifnum [1]{%
 \ifnum #1\expandafter \@firstoftwo
 \else \expandafter \@secondoftwo
 \fi
}%
\providecommand \@ifx [1]{%
 \ifx #1\expandafter \@firstoftwo
 \else \expandafter \@secondoftwo
 \fi
}%
\providecommand \natexlab [1]{#1}%
\providecommand \enquote  [1]{``#1''}%
\providecommand \bibnamefont  [1]{#1}%
\providecommand \bibfnamefont [1]{#1}%
\providecommand \citenamefont [1]{#1}%
\providecommand \href@noop [0]{\@secondoftwo}%
\providecommand \href [0]{\begingroup \@sanitize@url \@href}%
\providecommand \@href[1]{\@@startlink{#1}\@@href}%
\providecommand \@@href[1]{\endgroup#1\@@endlink}%
\providecommand \@sanitize@url [0]{\catcode `\\12\catcode `\$12\catcode
  `\&12\catcode `\#12\catcode `\^12\catcode `\_12\catcode `\%12\relax}%
\providecommand \@@startlink[1]{}%
\providecommand \@@endlink[0]{}%
\providecommand \url  [0]{\begingroup\@sanitize@url \@url }%
\providecommand \@url [1]{\endgroup\@href {#1}{\urlprefix }}%
\providecommand \urlprefix  [0]{URL }%
\providecommand \Eprint [0]{\href }%
\providecommand \doibase [0]{http://dx.doi.org/}%
\providecommand \selectlanguage [0]{\@gobble}%
\providecommand \bibinfo  [0]{\@secondoftwo}%
\providecommand \bibfield  [0]{\@secondoftwo}%
\providecommand \translation [1]{[#1]}%
\providecommand \BibitemOpen [0]{}%
\providecommand \bibitemStop [0]{}%
\providecommand \bibitemNoStop [0]{.\EOS\space}%
\providecommand \EOS [0]{\spacefactor3000\relax}%
\providecommand \BibitemShut  [1]{\csname bibitem#1\endcsname}%
\let\auto@bib@innerbib\@empty
%</preamble>
\bibitem [{\citenamefont {Amico}\ \emph {et~al.}(2008)\citenamefont {Amico},
  \citenamefont {Fazio}, \citenamefont {Osterloh},\ and\ \citenamefont
  {Vedral}}]{RevModPhys.80.517}%
  \BibitemOpen
  \bibfield  {author} {\bibinfo {author} {\bibfnamefont {L.}~\bibnamefont
  {Amico}}, \bibinfo {author} {\bibfnamefont {R.}~\bibnamefont {Fazio}},
  \bibinfo {author} {\bibfnamefont {A.}~\bibnamefont {Osterloh}}, \ and\
  \bibinfo {author} {\bibfnamefont {V.}~\bibnamefont {Vedral}},\ }\href
  {\doibase 10.1103/RevModPhys.80.517} {\bibfield  {journal} {\bibinfo
  {journal} {Rev. Mod. Phys.}\ }\textbf {\bibinfo {volume} {80}},\ \bibinfo
  {pages} {517} (\bibinfo {year} {2008})}\BibitemShut {NoStop}%
\bibitem [{\citenamefont {White}(1993)}]{PhysRevB.48.10345}%
  \BibitemOpen
  \bibfield  {author} {\bibinfo {author} {\bibfnamefont {S.~R.}\ \bibnamefont
  {White}},\ }\href {\doibase 10.1103/PhysRevB.48.10345} {\bibfield  {journal}
  {\bibinfo  {journal} {Phys. Rev. B}\ }\textbf {\bibinfo {volume} {48}},\
  \bibinfo {pages} {10345} (\bibinfo {year} {1993})}\BibitemShut {NoStop}%
\bibitem [{\citenamefont {Yang}\ and\ \citenamefont {Luo}(2023)}]{Yang_2023}%
  \BibitemOpen
  \bibfield  {author} {\bibinfo {author} {\bibfnamefont {Y.-T.}\ \bibnamefont
  {Yang}}\ and\ \bibinfo {author} {\bibfnamefont {H.-G.}\ \bibnamefont {Luo}},\
  }\href {\doibase 10.1088/0256-307X/40/2/020502} {\bibfield  {journal}
  {\bibinfo  {journal} {Chin. Phys. Lett.}\ }\textbf {\bibinfo {volume} {40}},\
  \bibinfo {pages} {020502} (\bibinfo {year} {2023})}\BibitemShut {NoStop}%
\bibitem [{\citenamefont {Jaeger}(1998)}]{Jaeger1998TheEC}%
  \BibitemOpen
  \bibfield  {author} {\bibinfo {author} {\bibfnamefont {G.}~\bibnamefont
  {Jaeger}},\ }\href {http://www.jstor.org/stable/41134053} {\bibfield
  {journal} {\bibinfo  {journal} {Arch. Hist. Exact Sci.}\ }\textbf {\bibinfo
  {volume} {53}},\ \bibinfo {pages} {51} (\bibinfo {year} {1998})}\BibitemShut
  {NoStop}%
\bibitem [{\citenamefont {Cejnar}\ \emph {et~al.}(2007)\citenamefont {Cejnar},
  \citenamefont {Heinze},\ and\ \citenamefont {Macek}}]{PhysRevLett.99.100601}%
  \BibitemOpen
  \bibfield  {author} {\bibinfo {author} {\bibfnamefont {P.}~\bibnamefont
  {Cejnar}}, \bibinfo {author} {\bibfnamefont {S.}~\bibnamefont {Heinze}}, \
  and\ \bibinfo {author} {\bibfnamefont {M.}~\bibnamefont {Macek}},\ }\href
  {\doibase 10.1103/PhysRevLett.99.100601} {\bibfield  {journal} {\bibinfo
  {journal} {Phys. Rev. Lett.}\ }\textbf {\bibinfo {volume} {99}},\ \bibinfo
  {pages} {100601} (\bibinfo {year} {2007})}\BibitemShut {NoStop}%
\bibitem [{\citenamefont {Arias}\ \emph {et~al.}(2003)\citenamefont {Arias},
  \citenamefont {Dukelsky},\ and\ \citenamefont
  {Garc\'{\i}a-Ramos}}]{PhysRevLett.91.162502}%
  \BibitemOpen
  \bibfield  {author} {\bibinfo {author} {\bibfnamefont {J.~M.}\ \bibnamefont
  {Arias}}, \bibinfo {author} {\bibfnamefont {J.}~\bibnamefont {Dukelsky}}, \
  and\ \bibinfo {author} {\bibfnamefont {J.~E.}\ \bibnamefont
  {Garc\'{\i}a-Ramos}},\ }\href {\doibase 10.1103/PhysRevLett.91.162502}
  {\bibfield  {journal} {\bibinfo  {journal} {Phys. Rev. Lett.}\ }\textbf
  {\bibinfo {volume} {91}},\ \bibinfo {pages} {162502} (\bibinfo {year}
  {2003})}\BibitemShut {NoStop}%
\bibitem [{\citenamefont {Sandvik}(2010)}]{PhysRevLett.104.137204}%
  \BibitemOpen
  \bibfield  {author} {\bibinfo {author} {\bibfnamefont {A.~W.}\ \bibnamefont
  {Sandvik}},\ }\href {\doibase 10.1103/PhysRevLett.104.137204} {\bibfield
  {journal} {\bibinfo  {journal} {Phys. Rev. Lett.}\ }\textbf {\bibinfo
  {volume} {104}},\ \bibinfo {pages} {137204} (\bibinfo {year}
  {2010})}\BibitemShut {NoStop}%
\bibitem [{\citenamefont {Wang}\ and\ \citenamefont
  {Sandvik}(2018)}]{PhysRevLett.121.107202}%
  \BibitemOpen
  \bibfield  {author} {\bibinfo {author} {\bibfnamefont {L.}~\bibnamefont
  {Wang}}\ and\ \bibinfo {author} {\bibfnamefont {A.~W.}\ \bibnamefont
  {Sandvik}},\ }\href {\doibase 10.1103/PhysRevLett.121.107202} {\bibfield
  {journal} {\bibinfo  {journal} {Phys. Rev. Lett.}\ }\textbf {\bibinfo
  {volume} {121}},\ \bibinfo {pages} {107202} (\bibinfo {year}
  {2018})}\BibitemShut {NoStop}%
\bibitem [{\citenamefont {Wang}\ \emph {et~al.}(2022)\citenamefont {Wang},
  \citenamefont {Zhang},\ and\ \citenamefont {Sandvik}}]{Wang_2022}%
  \BibitemOpen
  \bibfield  {author} {\bibinfo {author} {\bibfnamefont {L.}~\bibnamefont
  {Wang}}, \bibinfo {author} {\bibfnamefont {Y.}~\bibnamefont {Zhang}}, \ and\
  \bibinfo {author} {\bibfnamefont {A.~W.}\ \bibnamefont {Sandvik}},\ }\href
  {\doibase 10.1088/0256-307X/39/7/077502} {\bibfield  {journal} {\bibinfo
  {journal} {Chin. Phys. Lett.}\ }\textbf {\bibinfo {volume} {39}},\ \bibinfo
  {pages} {077502} (\bibinfo {year} {2022})}\BibitemShut {NoStop}%
\bibitem [{\citenamefont {Heyl}(2018)}]{heyl2018dynamical}%
  \BibitemOpen
  \bibfield  {author} {\bibinfo {author} {\bibfnamefont {M.}~\bibnamefont
  {Heyl}},\ }\href {\doibase 10.1088/1361-6633/aaaf9a} {\bibfield  {journal}
  {\bibinfo  {journal} {Rep. Prog. Phys}\ }\textbf {\bibinfo {volume} {81}},\
  \bibinfo {pages} {054001} (\bibinfo {year} {2018})}\BibitemShut {NoStop}%
\bibitem [{\citenamefont {Tian}\ \emph {et~al.}(2020)\citenamefont {Tian},
  \citenamefont {Yang}, \citenamefont {Qiu}, \citenamefont {Liang},
  \citenamefont {Yang}, \citenamefont {Xu},\ and\ \citenamefont
  {Duan}}]{PhysRevLett.124.043001}%
  \BibitemOpen
  \bibfield  {author} {\bibinfo {author} {\bibfnamefont {T.}~\bibnamefont
  {Tian}}, \bibinfo {author} {\bibfnamefont {H.-X.}\ \bibnamefont {Yang}},
  \bibinfo {author} {\bibfnamefont {L.-Y.}\ \bibnamefont {Qiu}}, \bibinfo
  {author} {\bibfnamefont {H.-Y.}\ \bibnamefont {Liang}}, \bibinfo {author}
  {\bibfnamefont {Y.-B.}\ \bibnamefont {Yang}}, \bibinfo {author}
  {\bibfnamefont {Y.}~\bibnamefont {Xu}}, \ and\ \bibinfo {author}
  {\bibfnamefont {L.-M.}\ \bibnamefont {Duan}},\ }\href {\doibase
  10.1103/PhysRevLett.124.043001} {\bibfield  {journal} {\bibinfo  {journal}
  {Phys. Rev. Lett.}\ }\textbf {\bibinfo {volume} {124}},\ \bibinfo {pages}
  {043001} (\bibinfo {year} {2020})}\BibitemShut {NoStop}%
\bibitem [{\citenamefont {P\'erez-Fern\'andez}\ \emph
  {et~al.}(2011)\citenamefont {P\'erez-Fern\'andez}, \citenamefont {Cejnar},
  \citenamefont {Arias}, \citenamefont {Dukelsky}, \citenamefont
  {Garc\'{\i}a-Ramos},\ and\ \citenamefont {Rela\~no}}]{PhysRevA.83.033802}%
  \BibitemOpen
  \bibfield  {author} {\bibinfo {author} {\bibfnamefont {P.}~\bibnamefont
  {P\'erez-Fern\'andez}}, \bibinfo {author} {\bibfnamefont {P.}~\bibnamefont
  {Cejnar}}, \bibinfo {author} {\bibfnamefont {J.~M.}\ \bibnamefont {Arias}},
  \bibinfo {author} {\bibfnamefont {J.}~\bibnamefont {Dukelsky}}, \bibinfo
  {author} {\bibfnamefont {J.~E.}\ \bibnamefont {Garc\'{\i}a-Ramos}}, \ and\
  \bibinfo {author} {\bibfnamefont {A.}~\bibnamefont {Rela\~no}},\ }\href
  {\doibase 10.1103/PhysRevA.83.033802} {\bibfield  {journal} {\bibinfo
  {journal} {Phys. Rev. A}\ }\textbf {\bibinfo {volume} {83}},\ \bibinfo
  {pages} {033802} (\bibinfo {year} {2011})}\BibitemShut {NoStop}%
\bibitem [{\citenamefont {Leyvraz}\ and\ \citenamefont
  {Heiss}(2005)}]{PhysRevLett.95.050402}%
  \BibitemOpen
  \bibfield  {author} {\bibinfo {author} {\bibfnamefont {F.}~\bibnamefont
  {Leyvraz}}\ and\ \bibinfo {author} {\bibfnamefont {W.~D.}\ \bibnamefont
  {Heiss}},\ }\href {\doibase 10.1103/PhysRevLett.95.050402} {\bibfield
  {journal} {\bibinfo  {journal} {Phys. Rev. Lett.}\ }\textbf {\bibinfo
  {volume} {95}},\ \bibinfo {pages} {050402} (\bibinfo {year}
  {2005})}\BibitemShut {NoStop}%
\bibitem [{\citenamefont {Heiss}\ \emph {et~al.}(2005)\citenamefont {Heiss},
  \citenamefont {Scholtz},\ and\ \citenamefont {Geyer}}]{Heiss_2005}%
  \BibitemOpen
  \bibfield  {author} {\bibinfo {author} {\bibfnamefont {W.~D.}\ \bibnamefont
  {Heiss}}, \bibinfo {author} {\bibfnamefont {F.~G.}\ \bibnamefont {Scholtz}},
  \ and\ \bibinfo {author} {\bibfnamefont {H.~B.}\ \bibnamefont {Geyer}},\
  }\href {\doibase 10.1088/0305-4470/38/9/002} {\bibfield  {journal} {\bibinfo
  {journal} {J. Phys. A Math. Theor.}\ }\textbf {\bibinfo {volume} {38}},\
  \bibinfo {pages} {1843} (\bibinfo {year} {2005})}\BibitemShut {NoStop}%
\bibitem [{\citenamefont {Santos}\ \emph {et~al.}(2016)\citenamefont {Santos},
  \citenamefont {T\'avora},\ and\ \citenamefont
  {P\'erez-Bernal}}]{PhysRevA.94.012113}%
  \BibitemOpen
  \bibfield  {author} {\bibinfo {author} {\bibfnamefont {L.~F.}\ \bibnamefont
  {Santos}}, \bibinfo {author} {\bibfnamefont {M.}~\bibnamefont {T\'avora}}, \
  and\ \bibinfo {author} {\bibfnamefont {F.}~\bibnamefont {P\'erez-Bernal}},\
  }\href {\doibase 10.1103/PhysRevA.94.012113} {\bibfield  {journal} {\bibinfo
  {journal} {Phys. Rev. A}\ }\textbf {\bibinfo {volume} {94}},\ \bibinfo
  {pages} {012113} (\bibinfo {year} {2016})}\BibitemShut {NoStop}%
\bibitem [{\citenamefont {Brandes}(2013)}]{PhysRevE.88.032133}%
  \BibitemOpen
  \bibfield  {author} {\bibinfo {author} {\bibfnamefont {T.}~\bibnamefont
  {Brandes}},\ }\href {\doibase 10.1103/PhysRevE.88.032133} {\bibfield
  {journal} {\bibinfo  {journal} {Phys. Rev. E}\ }\textbf {\bibinfo {volume}
  {88}},\ \bibinfo {pages} {032133} (\bibinfo {year} {2013})}\BibitemShut
  {NoStop}%
\bibitem [{\citenamefont {Puebla}\ \emph {et~al.}(2013)\citenamefont {Puebla},
  \citenamefont {Rela\~no},\ and\ \citenamefont
  {Retamosa}}]{PhysRevA.87.023819}%
  \BibitemOpen
  \bibfield  {author} {\bibinfo {author} {\bibfnamefont {R.}~\bibnamefont
  {Puebla}}, \bibinfo {author} {\bibfnamefont {A.}~\bibnamefont {Rela\~no}}, \
  and\ \bibinfo {author} {\bibfnamefont {J.}~\bibnamefont {Retamosa}},\ }\href
  {\doibase 10.1103/PhysRevA.87.023819} {\bibfield  {journal} {\bibinfo
  {journal} {Phys. Rev. A}\ }\textbf {\bibinfo {volume} {87}},\ \bibinfo
  {pages} {023819} (\bibinfo {year} {2013})}\BibitemShut {NoStop}%
\bibitem [{\citenamefont {P\'erez-Fern\'andez}\ \emph
  {et~al.}(2009)\citenamefont {P\'erez-Fern\'andez}, \citenamefont {Rela\~no},
  \citenamefont {Arias}, \citenamefont {Dukelsky},\ and\ \citenamefont
  {Garc\'{\i}a-Ramos}}]{PhysRevA.80.032111}%
  \BibitemOpen
  \bibfield  {author} {\bibinfo {author} {\bibfnamefont {P.}~\bibnamefont
  {P\'erez-Fern\'andez}}, \bibinfo {author} {\bibfnamefont {A.}~\bibnamefont
  {Rela\~no}}, \bibinfo {author} {\bibfnamefont {J.~M.}\ \bibnamefont {Arias}},
  \bibinfo {author} {\bibfnamefont {J.}~\bibnamefont {Dukelsky}}, \ and\
  \bibinfo {author} {\bibfnamefont {J.~E.}\ \bibnamefont {Garc\'{\i}a-Ramos}},\
  }\href {\doibase 10.1103/PhysRevA.80.032111} {\bibfield  {journal} {\bibinfo
  {journal} {Phys. Rev. A}\ }\textbf {\bibinfo {volume} {80}},\ \bibinfo
  {pages} {032111} (\bibinfo {year} {2009})}\BibitemShut {NoStop}%
\bibitem [{\citenamefont {Bastidas}\ \emph {et~al.}(2014)\citenamefont
  {Bastidas}, \citenamefont {P\'erez-Fern\'andez}, \citenamefont {Vogl},\ and\
  \citenamefont {Brandes}}]{PhysRevLett.112.140408}%
  \BibitemOpen
  \bibfield  {author} {\bibinfo {author} {\bibfnamefont {V.~M.}\ \bibnamefont
  {Bastidas}}, \bibinfo {author} {\bibfnamefont {P.}~\bibnamefont
  {P\'erez-Fern\'andez}}, \bibinfo {author} {\bibfnamefont {M.}~\bibnamefont
  {Vogl}}, \ and\ \bibinfo {author} {\bibfnamefont {T.}~\bibnamefont
  {Brandes}},\ }\href {\doibase 10.1103/PhysRevLett.112.140408} {\bibfield
  {journal} {\bibinfo  {journal} {Phys. Rev. Lett.}\ }\textbf {\bibinfo
  {volume} {112}},\ \bibinfo {pages} {140408} (\bibinfo {year}
  {2014})}\BibitemShut {NoStop}%
\bibitem [{\citenamefont {P\'erez-Bernal}\ and\ \citenamefont
  {\'Alvarez-Bajo}(2010)}]{PhysRevA.81.050101}%
  \BibitemOpen
  \bibfield  {author} {\bibinfo {author} {\bibfnamefont {F.}~\bibnamefont
  {P\'erez-Bernal}}\ and\ \bibinfo {author} {\bibfnamefont {O.}~\bibnamefont
  {\'Alvarez-Bajo}},\ }\href {\doibase 10.1103/PhysRevA.81.050101} {\bibfield
  {journal} {\bibinfo  {journal} {Phys. Rev. A}\ }\textbf {\bibinfo {volume}
  {81}},\ \bibinfo {pages} {050101} (\bibinfo {year} {2010})}\BibitemShut
  {NoStop}%
\bibitem [{\citenamefont {Khemani}\ \emph {et~al.}(2016)\citenamefont
  {Khemani}, \citenamefont {Pollmann},\ and\ \citenamefont
  {Sondhi}}]{PhysRevLett.116.247204}%
  \BibitemOpen
  \bibfield  {author} {\bibinfo {author} {\bibfnamefont {V.}~\bibnamefont
  {Khemani}}, \bibinfo {author} {\bibfnamefont {F.}~\bibnamefont {Pollmann}}, \
  and\ \bibinfo {author} {\bibfnamefont {S.~L.}\ \bibnamefont {Sondhi}},\
  }\href {\doibase 10.1103/PhysRevLett.116.247204} {\bibfield  {journal}
  {\bibinfo  {journal} {Phys. Rev. Lett.}\ }\textbf {\bibinfo {volume} {116}},\
  \bibinfo {pages} {247204} (\bibinfo {year} {2016})}\BibitemShut {NoStop}%
\bibitem [{\citenamefont {Yu}\ \emph {et~al.}(2017)\citenamefont {Yu},
  \citenamefont {Pekker},\ and\ \citenamefont
  {Clark}}]{PhysRevLett.118.017201}%
  \BibitemOpen
  \bibfield  {author} {\bibinfo {author} {\bibfnamefont {X.}~\bibnamefont
  {Yu}}, \bibinfo {author} {\bibfnamefont {D.}~\bibnamefont {Pekker}}, \ and\
  \bibinfo {author} {\bibfnamefont {B.~K.}\ \bibnamefont {Clark}},\ }\href
  {\doibase 10.1103/PhysRevLett.118.017201} {\bibfield  {journal} {\bibinfo
  {journal} {Phys. Rev. Lett.}\ }\textbf {\bibinfo {volume} {118}},\ \bibinfo
  {pages} {017201} (\bibinfo {year} {2017})}\BibitemShut {NoStop}%
\bibitem [{\citenamefont {Pancotti}\ \emph {et~al.}(2020)\citenamefont
  {Pancotti}, \citenamefont {Giudice}, \citenamefont {Cirac}, \citenamefont
  {Garrahan},\ and\ \citenamefont {Ba\~nuls}}]{PhysRevX.10.021051}%
  \BibitemOpen
  \bibfield  {author} {\bibinfo {author} {\bibfnamefont {N.}~\bibnamefont
  {Pancotti}}, \bibinfo {author} {\bibfnamefont {G.}~\bibnamefont {Giudice}},
  \bibinfo {author} {\bibfnamefont {J.~I.}\ \bibnamefont {Cirac}}, \bibinfo
  {author} {\bibfnamefont {J.~P.}\ \bibnamefont {Garrahan}}, \ and\ \bibinfo
  {author} {\bibfnamefont {M.~C.}\ \bibnamefont {Ba\~nuls}},\ }\href {\doibase
  10.1103/PhysRevX.10.021051} {\bibfield  {journal} {\bibinfo  {journal} {Phys.
  Rev. X}\ }\textbf {\bibinfo {volume} {10}},\ \bibinfo {pages} {021051}
  (\bibinfo {year} {2020})}\BibitemShut {NoStop}%
\bibitem [{\citenamefont {Alcaraz}\ \emph {et~al.}(2011)\citenamefont
  {Alcaraz}, \citenamefont {Berganza},\ and\ \citenamefont
  {Sierra}}]{PhysRevLett.106.201601}%
  \BibitemOpen
  \bibfield  {author} {\bibinfo {author} {\bibfnamefont {F.~C.}\ \bibnamefont
  {Alcaraz}}, \bibinfo {author} {\bibfnamefont {M.~I.~n.}\ \bibnamefont
  {Berganza}}, \ and\ \bibinfo {author} {\bibfnamefont {G.}~\bibnamefont
  {Sierra}},\ }\href {\doibase 10.1103/PhysRevLett.106.201601} {\bibfield
  {journal} {\bibinfo  {journal} {Phys. Rev. Lett.}\ }\textbf {\bibinfo
  {volume} {106}},\ \bibinfo {pages} {201601} (\bibinfo {year}
  {2011})}\BibitemShut {NoStop}%
\bibitem [{\citenamefont {\ifmmode \check{S}\else
  \v{S}\fi{}telmachovi\ifmmode~\check{c}\else \v{c}\fi{}}\ and\ \citenamefont
  {Bu\ifmmode~\check{z}\else \v{z}\fi{}ek}(2004)}]{PhysRevA.70.032313}%
  \BibitemOpen
  \bibfield  {author} {\bibinfo {author} {\bibfnamefont {P.}~\bibnamefont
  {\ifmmode \check{S}\else \v{S}\fi{}telmachovi\ifmmode~\check{c}\else
  \v{c}\fi{}}}\ and\ \bibinfo {author} {\bibfnamefont {V.}~\bibnamefont
  {Bu\ifmmode~\check{z}\else \v{z}\fi{}ek}},\ }\href {\doibase
  10.1103/PhysRevA.70.032313} {\bibfield  {journal} {\bibinfo  {journal} {Phys.
  Rev. A}\ }\textbf {\bibinfo {volume} {70}},\ \bibinfo {pages} {032313}
  (\bibinfo {year} {2004})}\BibitemShut {NoStop}%
\bibitem [{\citenamefont {Alba}\ \emph {et~al.}(2009)\citenamefont {Alba},
  \citenamefont {Fagotti},\ and\ \citenamefont {Calabrese}}]{Alba_2009}%
  \BibitemOpen
  \bibfield  {author} {\bibinfo {author} {\bibfnamefont {V.}~\bibnamefont
  {Alba}}, \bibinfo {author} {\bibfnamefont {M.}~\bibnamefont {Fagotti}}, \
  and\ \bibinfo {author} {\bibfnamefont {P.}~\bibnamefont {Calabrese}},\ }\href
  {\doibase 10.1088/1742-5468/2009/10/P10020} {\bibfield  {journal} {\bibinfo
  {journal} {J. Stat. Mech. Theory Exp}\ }\textbf {\bibinfo {volume} {2009}},\
  \bibinfo {pages} {P10020} (\bibinfo {year} {2009})}\BibitemShut {NoStop}%
\bibitem [{\citenamefont {Zou}\ \emph {et~al.}(2018)\citenamefont {Zou},
  \citenamefont {Milsted},\ and\ \citenamefont
  {Vidal}}]{PhysRevLett.121.230402}%
  \BibitemOpen
  \bibfield  {author} {\bibinfo {author} {\bibfnamefont {Y.}~\bibnamefont
  {Zou}}, \bibinfo {author} {\bibfnamefont {A.}~\bibnamefont {Milsted}}, \ and\
  \bibinfo {author} {\bibfnamefont {G.}~\bibnamefont {Vidal}},\ }\href
  {\doibase 10.1103/PhysRevLett.121.230402} {\bibfield  {journal} {\bibinfo
  {journal} {Phys. Rev. Lett.}\ }\textbf {\bibinfo {volume} {121}},\ \bibinfo
  {pages} {230402} (\bibinfo {year} {2018})}\BibitemShut {NoStop}%
\bibitem [{\citenamefont {Luo}\ \emph {et~al.}(2003)\citenamefont {Luo},
  \citenamefont {Xiang},\ and\ \citenamefont {Wang}}]{PhysRevLett.91.049701}%
  \BibitemOpen
  \bibfield  {author} {\bibinfo {author} {\bibfnamefont {H.~G.}\ \bibnamefont
  {Luo}}, \bibinfo {author} {\bibfnamefont {T.}~\bibnamefont {Xiang}}, \ and\
  \bibinfo {author} {\bibfnamefont {X.~Q.}\ \bibnamefont {Wang}},\ }\href
  {\doibase 10.1103/PhysRevLett.91.049701} {\bibfield  {journal} {\bibinfo
  {journal} {Phys. Rev. Lett.}\ }\textbf {\bibinfo {volume} {91}},\ \bibinfo
  {pages} {049701} (\bibinfo {year} {2003})}\BibitemShut {NoStop}%
\bibitem [{\citenamefont {Kj\"all}\ \emph {et~al.}(2014)\citenamefont
  {Kj\"all}, \citenamefont {Bardarson},\ and\ \citenamefont
  {Pollmann}}]{PhysRevLett.113.107204}%
  \BibitemOpen
  \bibfield  {author} {\bibinfo {author} {\bibfnamefont {J.~A.}\ \bibnamefont
  {Kj\"all}}, \bibinfo {author} {\bibfnamefont {J.~H.}\ \bibnamefont
  {Bardarson}}, \ and\ \bibinfo {author} {\bibfnamefont {F.}~\bibnamefont
  {Pollmann}},\ }\href {\doibase 10.1103/PhysRevLett.113.107204} {\bibfield
  {journal} {\bibinfo  {journal} {Phys. Rev. Lett.}\ }\textbf {\bibinfo
  {volume} {113}},\ \bibinfo {pages} {107204} (\bibinfo {year}
  {2014})}\BibitemShut {NoStop}%
\bibitem [{\citenamefont {Luitz}\ \emph {et~al.}(2015)\citenamefont {Luitz},
  \citenamefont {Laflorencie},\ and\ \citenamefont
  {Alet}}]{PhysRevB.91.081103}%
  \BibitemOpen
  \bibfield  {author} {\bibinfo {author} {\bibfnamefont {D.~J.}\ \bibnamefont
  {Luitz}}, \bibinfo {author} {\bibfnamefont {N.}~\bibnamefont {Laflorencie}},
  \ and\ \bibinfo {author} {\bibfnamefont {F.}~\bibnamefont {Alet}},\ }\href
  {\doibase 10.1103/PhysRevB.91.081103} {\bibfield  {journal} {\bibinfo
  {journal} {Phys. Rev. B}\ }\textbf {\bibinfo {volume} {91}},\ \bibinfo
  {pages} {081103} (\bibinfo {year} {2015})}\BibitemShut {NoStop}%
\bibitem [{\citenamefont {White}(1992)}]{PhysRevLett.69.2863}%
  \BibitemOpen
  \bibfield  {author} {\bibinfo {author} {\bibfnamefont {S.~R.}\ \bibnamefont
  {White}},\ }\href {\doibase 10.1103/PhysRevLett.69.2863} {\bibfield
  {journal} {\bibinfo  {journal} {Phys. Rev. Lett.}\ }\textbf {\bibinfo
  {volume} {69}},\ \bibinfo {pages} {2863} (\bibinfo {year}
  {1992})}\BibitemShut {NoStop}%
\bibitem [{\citenamefont {\"Ostlund}\ and\ \citenamefont
  {Rommer}(1995{\natexlab{a}})}]{PhysRevLett.75.3537}%
  \BibitemOpen
  \bibfield  {author} {\bibinfo {author} {\bibfnamefont {S.}~\bibnamefont
  {\"Ostlund}}\ and\ \bibinfo {author} {\bibfnamefont {S.}~\bibnamefont
  {Rommer}},\ }\href {\doibase 10.1103/PhysRevLett.75.3537} {\bibfield
  {journal} {\bibinfo  {journal} {Phys. Rev. Lett.}\ }\textbf {\bibinfo
  {volume} {75}},\ \bibinfo {pages} {3537} (\bibinfo {year}
  {1995}{\natexlab{a}})}\BibitemShut {NoStop}%
\bibitem [{\citenamefont {Schollw\"ock}(2005)}]{RevModPhys.77.259}%
  \BibitemOpen
  \bibfield  {author} {\bibinfo {author} {\bibfnamefont {U.}~\bibnamefont
  {Schollw\"ock}},\ }\href {\doibase 10.1103/RevModPhys.77.259} {\bibfield
  {journal} {\bibinfo  {journal} {Rev. Mod. Phys.}\ }\textbf {\bibinfo {volume}
  {77}},\ \bibinfo {pages} {259} (\bibinfo {year} {2005})}\BibitemShut
  {NoStop}%
\bibitem [{\citenamefont {Ba{\~{n}}uls}\ \emph {et~al.}(2013)\citenamefont
  {Ba{\~{n}}uls}, \citenamefont {Cichy}, \citenamefont {Cirac},\ and\
  \citenamefont {Jansen}}]{Bañuls2013}%
  \BibitemOpen
  \bibfield  {author} {\bibinfo {author} {\bibfnamefont {M.~C.}\ \bibnamefont
  {Ba{\~{n}}uls}}, \bibinfo {author} {\bibfnamefont {K.}~\bibnamefont {Cichy}},
  \bibinfo {author} {\bibfnamefont {J.~I.}\ \bibnamefont {Cirac}}, \ and\
  \bibinfo {author} {\bibfnamefont {K.}~\bibnamefont {Jansen}},\ }\href
  {\doibase 10.1007/JHEP11(2013)158} {\bibfield  {journal} {\bibinfo  {journal}
  {J. High Energy Phys.}\ }\textbf {\bibinfo {volume} {2013}},\ \bibinfo
  {pages} {158} (\bibinfo {year} {2013})}\BibitemShut {NoStop}%
\bibitem [{\citenamefont {Vidal}(2003)}]{PhysRevLett.91.147902}%
  \BibitemOpen
  \bibfield  {author} {\bibinfo {author} {\bibfnamefont {G.}~\bibnamefont
  {Vidal}},\ }\href {\doibase 10.1103/PhysRevLett.91.147902} {\bibfield
  {journal} {\bibinfo  {journal} {Phys. Rev. Lett.}\ }\textbf {\bibinfo
  {volume} {91}},\ \bibinfo {pages} {147902} (\bibinfo {year}
  {2003})}\BibitemShut {NoStop}%
\bibitem [{\citenamefont {White}\ and\ \citenamefont
  {Affleck}(2008)}]{PhysRevB.77.134437}%
  \BibitemOpen
  \bibfield  {author} {\bibinfo {author} {\bibfnamefont {S.~R.}\ \bibnamefont
  {White}}\ and\ \bibinfo {author} {\bibfnamefont {I.}~\bibnamefont
  {Affleck}},\ }\href {\doibase 10.1103/PhysRevB.77.134437} {\bibfield
  {journal} {\bibinfo  {journal} {Phys. Rev. B}\ }\textbf {\bibinfo {volume}
  {77}},\ \bibinfo {pages} {134437} (\bibinfo {year} {2008})}\BibitemShut
  {NoStop}%
\bibitem [{\citenamefont {\"Ostlund}\ and\ \citenamefont
  {Rommer}(1995{\natexlab{b}})}]{Ostlund1995}%
  \BibitemOpen
  \bibfield  {author} {\bibinfo {author} {\bibfnamefont {S.}~\bibnamefont
  {\"Ostlund}}\ and\ \bibinfo {author} {\bibfnamefont {S.}~\bibnamefont
  {Rommer}},\ }\href {\doibase 10.1103/PhysRevLett.75.3537} {\bibfield
  {journal} {\bibinfo  {journal} {Phys. Rev. Lett.}\ }\textbf {\bibinfo
  {volume} {75}},\ \bibinfo {pages} {3537} (\bibinfo {year}
  {1995}{\natexlab{b}})}\BibitemShut {NoStop}%
\bibitem [{\citenamefont {Haegeman}\ \emph {et~al.}(2012)\citenamefont
  {Haegeman}, \citenamefont {Pirvu}, \citenamefont {Weir}, \citenamefont
  {Cirac}, \citenamefont {Osborne}, \citenamefont {Verschelde},\ and\
  \citenamefont {Verstraete}}]{Haegeman2012}%
  \BibitemOpen
  \bibfield  {author} {\bibinfo {author} {\bibfnamefont {J.}~\bibnamefont
  {Haegeman}}, \bibinfo {author} {\bibfnamefont {B.}~\bibnamefont {Pirvu}},
  \bibinfo {author} {\bibfnamefont {D.~J.}\ \bibnamefont {Weir}}, \bibinfo
  {author} {\bibfnamefont {J.~I.}\ \bibnamefont {Cirac}}, \bibinfo {author}
  {\bibfnamefont {T.~J.}\ \bibnamefont {Osborne}}, \bibinfo {author}
  {\bibfnamefont {H.}~\bibnamefont {Verschelde}}, \ and\ \bibinfo {author}
  {\bibfnamefont {F.}~\bibnamefont {Verstraete}},\ }\href {\doibase
  10.1103/PhysRevB.85.100408} {\bibfield  {journal} {\bibinfo  {journal} {Phys.
  Rev. B}\ }\textbf {\bibinfo {volume} {85}},\ \bibinfo {pages} {100408}
  (\bibinfo {year} {2012})}\BibitemShut {NoStop}%
\bibitem [{\citenamefont {Vanderstraeten}\ \emph {et~al.}(2015)\citenamefont
  {Vanderstraeten}, \citenamefont {Mariën}, \citenamefont {Verstraete},\ and\
  \citenamefont {Haegeman}}]{Vanderstraeten2015}%
  \BibitemOpen
  \bibfield  {author} {\bibinfo {author} {\bibfnamefont {L.}~\bibnamefont
  {Vanderstraeten}}, \bibinfo {author} {\bibfnamefont {M.}~\bibnamefont
  {Mariën}}, \bibinfo {author} {\bibfnamefont {F.}~\bibnamefont {Verstraete}},
  \ and\ \bibinfo {author} {\bibfnamefont {J.}~\bibnamefont {Haegeman}},\
  }\href {\doibase 10.1103/physrevb.92.201111} {\bibfield  {journal} {\bibinfo
  {journal} {Physical Review B}\ }\textbf {\bibinfo {volume} {92}},\ \bibinfo
  {pages} {201111} (\bibinfo {year} {2015})}\BibitemShut {NoStop}%
\bibitem [{\citenamefont {Vanderstraeten}\ \emph {et~al.}(2019)\citenamefont
  {Vanderstraeten}, \citenamefont {Haegeman},\ and\ \citenamefont
  {Verstraete}}]{Vanderstraeten2019}%
  \BibitemOpen
  \bibfield  {author} {\bibinfo {author} {\bibfnamefont {L.}~\bibnamefont
  {Vanderstraeten}}, \bibinfo {author} {\bibfnamefont {J.}~\bibnamefont
  {Haegeman}}, \ and\ \bibinfo {author} {\bibfnamefont {F.}~\bibnamefont
  {Verstraete}},\ }\href {\doibase 10.1103/physrevb.99.165121} {\bibfield
  {journal} {\bibinfo  {journal} {Physical Review B}\ }\textbf {\bibinfo
  {volume} {99}},\ \bibinfo {pages} {165121} (\bibinfo {year}
  {2019})}\BibitemShut {NoStop}%
\bibitem [{\citenamefont {Ponsioen}\ and\ \citenamefont
  {Corboz}(2020)}]{Ponsioen2020}%
  \BibitemOpen
  \bibfield  {author} {\bibinfo {author} {\bibfnamefont {B.}~\bibnamefont
  {Ponsioen}}\ and\ \bibinfo {author} {\bibfnamefont {P.}~\bibnamefont
  {Corboz}},\ }\href {\doibase 10.1103/physrevb.101.195109} {\bibfield
  {journal} {\bibinfo  {journal} {Physical Review B}\ }\textbf {\bibinfo
  {volume} {101}},\ \bibinfo {pages} {195109} (\bibinfo {year}
  {2020})}\BibitemShut {NoStop}%
\bibitem [{\citenamefont {Ponsioen}\ \emph {et~al.}(2022)\citenamefont
  {Ponsioen}, \citenamefont {Assaad},\ and\ \citenamefont
  {Corboz}}]{Ponsioen2022}%
  \BibitemOpen
  \bibfield  {author} {\bibinfo {author} {\bibfnamefont {B.}~\bibnamefont
  {Ponsioen}}, \bibinfo {author} {\bibfnamefont {F.~F.}\ \bibnamefont
  {Assaad}}, \ and\ \bibinfo {author} {\bibfnamefont {P.}~\bibnamefont
  {Corboz}},\ }\href {\doibase 10.21468/SciPostPhys.12.1.006} {\bibfield
  {journal} {\bibinfo  {journal} {SciPost Phys.}\ }\textbf {\bibinfo {volume}
  {12}},\ \bibinfo {pages} {6} (\bibinfo {year} {2022})}\BibitemShut {NoStop}%
\bibitem [{\citenamefont {Chi}\ \emph {et~al.}(2022)\citenamefont {Chi},
  \citenamefont {Liu}, \citenamefont {Wan}, \citenamefont {Liao},\ and\
  \citenamefont {Xiang}}]{Chi2022}%
  \BibitemOpen
  \bibfield  {author} {\bibinfo {author} {\bibfnamefont {R.}~\bibnamefont
  {Chi}}, \bibinfo {author} {\bibfnamefont {Y.}~\bibnamefont {Liu}}, \bibinfo
  {author} {\bibfnamefont {Y.}~\bibnamefont {Wan}}, \bibinfo {author}
  {\bibfnamefont {H.-J.}\ \bibnamefont {Liao}}, \ and\ \bibinfo {author}
  {\bibfnamefont {T.}~\bibnamefont {Xiang}},\ }\href {\doibase
  10.1103/PhysRevLett.129.227201} {\bibfield  {journal} {\bibinfo  {journal}
  {Phys. Rev. Lett.}\ }\textbf {\bibinfo {volume} {129}},\ \bibinfo {pages}
  {227201} (\bibinfo {year} {2022})}\BibitemShut {NoStop}%
\bibitem [{\citenamefont {Jiang}\ \emph {et~al.}(2008)\citenamefont {Jiang},
  \citenamefont {Weng},\ and\ \citenamefont {Xiang}}]{PhysRevLett.101.090603}%
  \BibitemOpen
  \bibfield  {author} {\bibinfo {author} {\bibfnamefont {H.~C.}\ \bibnamefont
  {Jiang}}, \bibinfo {author} {\bibfnamefont {Z.~Y.}\ \bibnamefont {Weng}}, \
  and\ \bibinfo {author} {\bibfnamefont {T.}~\bibnamefont {Xiang}},\ }\href
  {\doibase 10.1103/PhysRevLett.101.090603} {\bibfield  {journal} {\bibinfo
  {journal} {Phys. Rev. Lett.}\ }\textbf {\bibinfo {volume} {101}},\ \bibinfo
  {pages} {090603} (\bibinfo {year} {2008})}\BibitemShut {NoStop}%
\bibitem [{\citenamefont {Hauru}\ \emph {et~al.}(2021)\citenamefont {Hauru},
  \citenamefont {Damme},\ and\ \citenamefont
  {Haegeman}}]{10.21468/SciPostPhys.10.2.040}%
  \BibitemOpen
  \bibfield  {author} {\bibinfo {author} {\bibfnamefont {M.}~\bibnamefont
  {Hauru}}, \bibinfo {author} {\bibfnamefont {M.~V.}\ \bibnamefont {Damme}}, \
  and\ \bibinfo {author} {\bibfnamefont {J.}~\bibnamefont {Haegeman}},\ }\href
  {\doibase 10.21468/SciPostPhys.10.2.040} {\bibfield  {journal} {\bibinfo
  {journal} {SciPost Phys.}\ }\textbf {\bibinfo {volume} {10}},\ \bibinfo
  {pages} {040} (\bibinfo {year} {2021})}\BibitemShut {NoStop}%
\bibitem [{\citenamefont {Ring}\ and\ \citenamefont {Wirth}(2012)}]{RingOpti}%
  \BibitemOpen
  \bibfield  {author} {\bibinfo {author} {\bibfnamefont {W.}~\bibnamefont
  {Ring}}\ and\ \bibinfo {author} {\bibfnamefont {B.}~\bibnamefont {Wirth}},\
  }\href {\doibase 10.1137/11082885X} {\bibfield  {journal} {\bibinfo
  {journal} {SIAM J. Optim.}\ }\textbf {\bibinfo {volume} {22}},\ \bibinfo
  {pages} {596} (\bibinfo {year} {2012})}\BibitemShut {NoStop}%
\bibitem [{\citenamefont {Huang}\ \emph {et~al.}(2018)\citenamefont {Huang},
  \citenamefont {Liao}, \citenamefont {Liu}, \citenamefont {Xie}, \citenamefont
  {Xie}, \citenamefont {Zhao}, \citenamefont {Chen},\ and\ \citenamefont
  {Xiang}}]{Huang_2018}%
  \BibitemOpen
  \bibfield  {author} {\bibinfo {author} {\bibfnamefont {R.-Z.}\ \bibnamefont
  {Huang}}, \bibinfo {author} {\bibfnamefont {H.-J.}\ \bibnamefont {Liao}},
  \bibinfo {author} {\bibfnamefont {Z.-Y.}\ \bibnamefont {Liu}}, \bibinfo
  {author} {\bibfnamefont {H.-D.}\ \bibnamefont {Xie}}, \bibinfo {author}
  {\bibfnamefont {Z.-Y.}\ \bibnamefont {Xie}}, \bibinfo {author} {\bibfnamefont
  {H.-H.}\ \bibnamefont {Zhao}}, \bibinfo {author} {\bibfnamefont
  {J.}~\bibnamefont {Chen}}, \ and\ \bibinfo {author} {\bibfnamefont
  {T.}~\bibnamefont {Xiang}},\ }\href {\doibase 10.1088/1674-1056/27/7/070501}
  {\bibfield  {journal} {\bibinfo  {journal} {Chin. Phys. B}\ }\textbf
  {\bibinfo {volume} {27}},\ \bibinfo {pages} {070501} (\bibinfo {year}
  {2018})}\BibitemShut {NoStop}%
\bibitem [{\citenamefont {Xiang}(ress)}]{Xiang2023}%
  \BibitemOpen
  \bibfield  {author} {\bibinfo {author} {\bibfnamefont {T.}~\bibnamefont
  {Xiang}},\ }\href@noop {} {\emph {\bibinfo {title} {{Density Matrix and
  Tensor Network Renormalization}}}}\ (\bibinfo  {publisher} {Cambridge
  University Press},\ \bibinfo {year} {2023, in press})\BibitemShut {NoStop}%
\bibitem [{\citenamefont {{Baker}}\ \emph {et~al.}(2021)\citenamefont
  {{Baker}}, \citenamefont {{Foley}},\ and\ \citenamefont
  {{S{\'e}n{\'e}chal}}}]{senechal2022direct}%
  \BibitemOpen
  \bibfield  {author} {\bibinfo {author} {\bibfnamefont {T.~E.}\ \bibnamefont
  {{Baker}}}, \bibinfo {author} {\bibfnamefont {A.}~\bibnamefont {{Foley}}}, \
  and\ \bibinfo {author} {\bibfnamefont {D.}~\bibnamefont
  {{S{\'e}n{\'e}chal}}},\ }\href
  {https://ui.adsabs.harvard.edu/abs/2021arXiv210908181B} {\bibfield  {journal}
  {\bibinfo  {journal} {arXiv:2109.08181}\ } (\bibinfo {year}
  {2021})}\BibitemShut {NoStop}%
\bibitem [{\citenamefont {Verstraete}\ \emph {et~al.}(2004)\citenamefont
  {Verstraete}, \citenamefont {Porras},\ and\ \citenamefont
  {Cirac}}]{PhysRevLett.93.227205}%
  \BibitemOpen
  \bibfield  {author} {\bibinfo {author} {\bibfnamefont {F.}~\bibnamefont
  {Verstraete}}, \bibinfo {author} {\bibfnamefont {D.}~\bibnamefont {Porras}},
  \ and\ \bibinfo {author} {\bibfnamefont {J.~I.}\ \bibnamefont {Cirac}},\
  }\href {\doibase 10.1103/PhysRevLett.93.227205} {\bibfield  {journal}
  {\bibinfo  {journal} {Phys. Rev. Lett.}\ }\textbf {\bibinfo {volume} {93}},\
  \bibinfo {pages} {227205} (\bibinfo {year} {2004})}\BibitemShut {NoStop}%
\bibitem [{\citenamefont {Liao}\ \emph {et~al.}(2019)\citenamefont {Liao},
  \citenamefont {Liu}, \citenamefont {Wang},\ and\ \citenamefont
  {Xiang}}]{PhysRevX.9.031041}%
  \BibitemOpen
  \bibfield  {author} {\bibinfo {author} {\bibfnamefont {H.-J.}\ \bibnamefont
  {Liao}}, \bibinfo {author} {\bibfnamefont {J.-G.}\ \bibnamefont {Liu}},
  \bibinfo {author} {\bibfnamefont {L.}~\bibnamefont {Wang}}, \ and\ \bibinfo
  {author} {\bibfnamefont {T.}~\bibnamefont {Xiang}},\ }\href {\doibase
  10.1103/PhysRevX.9.031041} {\bibfield  {journal} {\bibinfo  {journal} {Phys.
  Rev. X}\ }\textbf {\bibinfo {volume} {9}},\ \bibinfo {pages} {031041}
  (\bibinfo {year} {2019})}\BibitemShut {NoStop}%
\bibitem [{\citenamefont {Gross}\ \emph {et~al.}(1988)\citenamefont {Gross},
  \citenamefont {Oliveira},\ and\ \citenamefont {Kohn}}]{PhysRevA.37.2805}%
  \BibitemOpen
  \bibfield  {author} {\bibinfo {author} {\bibfnamefont {E.~K.~U.}\
  \bibnamefont {Gross}}, \bibinfo {author} {\bibfnamefont {L.~N.}\ \bibnamefont
  {Oliveira}}, \ and\ \bibinfo {author} {\bibfnamefont {W.}~\bibnamefont
  {Kohn}},\ }\href {\doibase 10.1103/PhysRevA.37.2805} {\bibfield  {journal}
  {\bibinfo  {journal} {Phys. Rev. A}\ }\textbf {\bibinfo {volume} {37}},\
  \bibinfo {pages} {2805} (\bibinfo {year} {1988})}\BibitemShut {NoStop}%
\bibitem [{\citenamefont {Sato}\ and\ \citenamefont
  {Iwai}(2013)}]{sato2013riemannian}%
  \BibitemOpen
  \bibfield  {author} {\bibinfo {author} {\bibfnamefont {H.}~\bibnamefont
  {Sato}}\ and\ \bibinfo {author} {\bibfnamefont {T.}~\bibnamefont {Iwai}},\
  }\href {\doibase 10.1137/120872887} {\bibfield  {journal} {\bibinfo
  {journal} {SIAM J. Optim.}\ }\textbf {\bibinfo {volume} {23}},\ \bibinfo
  {pages} {188} (\bibinfo {year} {2013})}\BibitemShut {NoStop}%
\bibitem [{\citenamefont {Wei}\ \emph {et~al.}(2022)\citenamefont {Wei},
  \citenamefont {Shen}, \citenamefont {Sun}, \citenamefont {Gao},\ and\
  \citenamefont {Ren}}]{wei2022neighborhood}%
  \BibitemOpen
  \bibfield  {author} {\bibinfo {author} {\bibfnamefont {D.}~\bibnamefont
  {Wei}}, \bibinfo {author} {\bibfnamefont {X.}~\bibnamefont {Shen}}, \bibinfo
  {author} {\bibfnamefont {Q.}~\bibnamefont {Sun}}, \bibinfo {author}
  {\bibfnamefont {X.}~\bibnamefont {Gao}}, \ and\ \bibinfo {author}
  {\bibfnamefont {Z.}~\bibnamefont {Ren}},\ }\href {\doibase
  https://doi.org/10.1016/j.patcog.2021.108335} {\bibfield  {journal} {\bibinfo
   {journal} {Pattern Recognit.}\ }\textbf {\bibinfo {volume} {122}},\ \bibinfo
  {pages} {108335} (\bibinfo {year} {2022})}\BibitemShut {NoStop}%
\bibitem [{\citenamefont {{Cui}}\ \emph {et~al.}(2022)\citenamefont {{Cui}},
  \citenamefont {{Li}}, \citenamefont {{Dong}},\ and\ \citenamefont
  {{Liu}}}]{cui2022taotf}%
  \BibitemOpen
  \bibfield  {author} {\bibinfo {author} {\bibfnamefont {T.}~\bibnamefont
  {{Cui}}}, \bibinfo {author} {\bibfnamefont {J.}~\bibnamefont {{Li}}},
  \bibinfo {author} {\bibfnamefont {Y.}~\bibnamefont {{Dong}}}, \ and\ \bibinfo
  {author} {\bibfnamefont {L.}~\bibnamefont {{Liu}}},\ }\href
  {https://ui.adsabs.harvard.edu/abs/2022arXiv221113902C} {\bibfield  {journal}
  {\bibinfo  {journal} {arXiv:2211.13902}\ } (\bibinfo {year}
  {2022})}\BibitemShut {NoStop}%
\bibitem [{\citenamefont {Wen}\ \emph {et~al.}(2016)\citenamefont {Wen},
  \citenamefont {Yang}, \citenamefont {Liu},\ and\ \citenamefont
  {Zhang}}]{Wen2016}%
  \BibitemOpen
  \bibfield  {author} {\bibinfo {author} {\bibfnamefont {Z.}~\bibnamefont
  {Wen}}, \bibinfo {author} {\bibfnamefont {C.}~\bibnamefont {Yang}}, \bibinfo
  {author} {\bibfnamefont {X.}~\bibnamefont {Liu}}, \ and\ \bibinfo {author}
  {\bibfnamefont {Y.}~\bibnamefont {Zhang}},\ }\href {\doibase
  10.1007/s10915-015-0061-0} {\bibfield  {journal} {\bibinfo  {journal} {J.
  Sci. Comput.}\ }\textbf {\bibinfo {volume} {66}},\ \bibinfo {pages} {1175}
  (\bibinfo {year} {2016})}\BibitemShut {NoStop}%
\bibitem [{\citenamefont {Saad}(2011)}]{saad2011numerical}%
  \BibitemOpen
  \bibfield  {author} {\bibinfo {author} {\bibfnamefont {Y.}~\bibnamefont
  {Saad}},\ }\href@noop {} {\emph {\bibinfo {title} {Numerical methods for
  large eigenvalue problems: revised edition}}}\ (\bibinfo  {publisher}
  {SIAM},\ \bibinfo {year} {2011})\BibitemShut {NoStop}%
\bibitem [{\citenamefont {Zhang}\ \emph {et~al.}(2014)\citenamefont {Zhang},
  \citenamefont {Zhu}, \citenamefont {Wen},\ and\ \citenamefont
  {Zhou}}]{zhang2014gradient}%
  \BibitemOpen
  \bibfield  {author} {\bibinfo {author} {\bibfnamefont {X.}~\bibnamefont
  {Zhang}}, \bibinfo {author} {\bibfnamefont {J.}~\bibnamefont {Zhu}}, \bibinfo
  {author} {\bibfnamefont {Z.}~\bibnamefont {Wen}}, \ and\ \bibinfo {author}
  {\bibfnamefont {A.}~\bibnamefont {Zhou}},\ }\href {\doibase
  10.1137/130932934} {\bibfield  {journal} {\bibinfo  {journal} {SIAM. J. Sci.
  Comput.}\ }\textbf {\bibinfo {volume} {36}},\ \bibinfo {pages} {C265}
  (\bibinfo {year} {2014})}\BibitemShut {NoStop}%
\bibitem [{\citenamefont {Altmann}\ \emph {et~al.}(2022)\citenamefont
  {Altmann}, \citenamefont {Peterseim},\ and\ \citenamefont
  {Stykel}}]{altmann2022energy}%
  \BibitemOpen
  \bibfield  {author} {\bibinfo {author} {\bibfnamefont {R.}~\bibnamefont
  {Altmann}}, \bibinfo {author} {\bibfnamefont {D.}~\bibnamefont {Peterseim}},
  \ and\ \bibinfo {author} {\bibfnamefont {T.}~\bibnamefont {Stykel}},\ }\href
  {https://doi.org/10.1051/m2an/2022036} {\bibfield  {journal} {\bibinfo
  {journal} {ESAIM Math. Model. Numer. Anal.}\ }\textbf {\bibinfo {volume}
  {56}} (\bibinfo {year} {2022})}\BibitemShut {NoStop}%
\bibitem [{\citenamefont {Vidal}(2007)}]{PhysRevLett.99.220405}%
  \BibitemOpen
  \bibfield  {author} {\bibinfo {author} {\bibfnamefont {G.}~\bibnamefont
  {Vidal}},\ }\href {\doibase 10.1103/PhysRevLett.99.220405} {\bibfield
  {journal} {\bibinfo  {journal} {Phys. Rev. Lett.}\ }\textbf {\bibinfo
  {volume} {99}},\ \bibinfo {pages} {220405} (\bibinfo {year}
  {2007})}\BibitemShut {NoStop}%
\bibitem [{\citenamefont {Luchnikov}\ \emph {et~al.}(2021)\citenamefont
  {Luchnikov}, \citenamefont {Krechetov},\ and\ \citenamefont
  {Filippov}}]{Luchnikov_2021}%
  \BibitemOpen
  \bibfield  {author} {\bibinfo {author} {\bibfnamefont {I.~A.}\ \bibnamefont
  {Luchnikov}}, \bibinfo {author} {\bibfnamefont {M.~E.}\ \bibnamefont
  {Krechetov}}, \ and\ \bibinfo {author} {\bibfnamefont {S.~N.}\ \bibnamefont
  {Filippov}},\ }\href {\doibase 10.1088/1367-2630/ac0b02} {\bibfield
  {journal} {\bibinfo  {journal} {New J. Phys.}\ }\textbf {\bibinfo {volume}
  {23}},\ \bibinfo {pages} {073006} (\bibinfo {year} {2021})}\BibitemShut
  {NoStop}%
\bibitem [{\citenamefont {Zaletel}\ and\ \citenamefont
  {Pollmann}(2020)}]{PhysRevLett.124.037201}%
  \BibitemOpen
  \bibfield  {author} {\bibinfo {author} {\bibfnamefont {M.~P.}\ \bibnamefont
  {Zaletel}}\ and\ \bibinfo {author} {\bibfnamefont {F.}~\bibnamefont
  {Pollmann}},\ }\href {\doibase 10.1103/PhysRevLett.124.037201} {\bibfield
  {journal} {\bibinfo  {journal} {Phys. Rev. Lett.}\ }\textbf {\bibinfo
  {volume} {124}},\ \bibinfo {pages} {037201} (\bibinfo {year}
  {2020})}\BibitemShut {NoStop}%
\bibitem [{\citenamefont {Tepaske}\ and\ \citenamefont
  {Luitz}(2021)}]{PhysRevResearch.3.023236}%
  \BibitemOpen
  \bibfield  {author} {\bibinfo {author} {\bibfnamefont {M.~S.~J.}\
  \bibnamefont {Tepaske}}\ and\ \bibinfo {author} {\bibfnamefont {D.~J.}\
  \bibnamefont {Luitz}},\ }\href {\doibase 10.1103/PhysRevResearch.3.023236}
  {\bibfield  {journal} {\bibinfo  {journal} {Phys. Rev. Res.}\ }\textbf
  {\bibinfo {volume} {3}},\ \bibinfo {pages} {023236} (\bibinfo {year}
  {2021})}\BibitemShut {NoStop}%
\bibitem [{\citenamefont {Absil}\ \emph {et~al.}(2008)\citenamefont {Absil},
  \citenamefont {Mahony},\ and\ \citenamefont
  {Sepulchre}}]{absil2008optimization}%
  \BibitemOpen
  \bibfield  {author} {\bibinfo {author} {\bibfnamefont {P.-A.}\ \bibnamefont
  {Absil}}, \bibinfo {author} {\bibfnamefont {R.}~\bibnamefont {Mahony}}, \
  and\ \bibinfo {author} {\bibfnamefont {R.}~\bibnamefont {Sepulchre}},\
  }\href@noop {} {\emph {\bibinfo {title} {Optimization algorithms on matrix
  manifolds}}}\ (\bibinfo  {publisher} {Princeton University Press},\ \bibinfo
  {year} {2008})\BibitemShut {NoStop}%
\bibitem [{\citenamefont {Zhu}(2017)}]{Zhu2017}%
  \BibitemOpen
  \bibfield  {author} {\bibinfo {author} {\bibfnamefont {X.}~\bibnamefont
  {Zhu}},\ }\href {\doibase 10.1007/s10589-016-9883-4} {\bibfield  {journal}
  {\bibinfo  {journal} {Comput. Optim. Appl.}\ }\textbf {\bibinfo {volume}
  {67}},\ \bibinfo {pages} {73} (\bibinfo {year} {2017})}\BibitemShut {NoStop}%
\bibitem [{\citenamefont {Zhang}\ \emph {et~al.}(2016)\citenamefont {Zhang},
  \citenamefont {J.~Reddi},\ and\ \citenamefont {Sra}}]{zhang2016riemannian}%
  \BibitemOpen
  \bibfield  {author} {\bibinfo {author} {\bibfnamefont {H.}~\bibnamefont
  {Zhang}}, \bibinfo {author} {\bibfnamefont {S.}~\bibnamefont {J.~Reddi}}, \
  and\ \bibinfo {author} {\bibfnamefont {S.}~\bibnamefont {Sra}},\ }\bibfield
  {booktitle} {\emph {\bibinfo {booktitle} {Advances in Neural Information
  Processing Systems}},\ }\href
  {https://proceedings.neurips.cc/paper_files/paper/2016/file/98e6f17209029f4ae6dc9d88ec8eac2c-Paper.pdf}
  {\ \textbf {\bibinfo {volume} {29}} (\bibinfo {year} {2016})}\BibitemShut
  {NoStop}%
\bibitem [{\citenamefont {Hu}\ \emph {et~al.}(2018)\citenamefont {Hu},
  \citenamefont {Milzarek}, \citenamefont {Wen},\ and\ \citenamefont
  {Yuan}}]{hu2018adaptive}%
  \BibitemOpen
  \bibfield  {author} {\bibinfo {author} {\bibfnamefont {J.}~\bibnamefont
  {Hu}}, \bibinfo {author} {\bibfnamefont {A.}~\bibnamefont {Milzarek}},
  \bibinfo {author} {\bibfnamefont {Z.}~\bibnamefont {Wen}}, \ and\ \bibinfo
  {author} {\bibfnamefont {Y.}~\bibnamefont {Yuan}},\ }\href {\doibase
  10.1137/17M1142478} {\bibfield  {journal} {\bibinfo  {journal} {SIAM J.
  Matrix Anal. Appl.}\ }\textbf {\bibinfo {volume} {39}},\ \bibinfo {pages}
  {1181} (\bibinfo {year} {2018})}\BibitemShut {NoStop}%
\bibitem [{\citenamefont {Nishimori}\ and\ \citenamefont
  {Akaho}(2005)}]{nishimori2005learning}%
  \BibitemOpen
  \bibfield  {author} {\bibinfo {author} {\bibfnamefont {Y.}~\bibnamefont
  {Nishimori}}\ and\ \bibinfo {author} {\bibfnamefont {S.}~\bibnamefont
  {Akaho}},\ }\href {\doibase https://doi.org/10.1016/j.neucom.2004.11.035}
  {\bibfield  {journal} {\bibinfo  {journal} {Neurocomputing}\ }\textbf
  {\bibinfo {volume} {67}},\ \bibinfo {pages} {106} (\bibinfo {year}
  {2005})}\BibitemShut {NoStop}%
\bibitem [{\citenamefont {Wen}\ and\ \citenamefont {Yin}(2013)}]{Wen2013}%
  \BibitemOpen
  \bibfield  {author} {\bibinfo {author} {\bibfnamefont {Z.}~\bibnamefont
  {Wen}}\ and\ \bibinfo {author} {\bibfnamefont {W.}~\bibnamefont {Yin}},\
  }\href {\doibase 10.1007/s10107-012-0584-1} {\bibfield  {journal} {\bibinfo
  {journal} {Math. Program.}\ }\textbf {\bibinfo {volume} {142}},\ \bibinfo
  {pages} {397} (\bibinfo {year} {2013})}\BibitemShut {NoStop}%
\bibitem [{\citenamefont {Zhu}\ and\ \citenamefont
  {Sato}(2021)}]{zhu2021cayley}%
  \BibitemOpen
  \bibfield  {author} {\bibinfo {author} {\bibfnamefont {X.}~\bibnamefont
  {Zhu}}\ and\ \bibinfo {author} {\bibfnamefont {H.}~\bibnamefont {Sato}},\
  }\href {\doibase 10.1007/s10444-021-09880-9} {\bibfield  {journal} {\bibinfo
  {journal} {Adv. Comput. Math.}\ }\textbf {\bibinfo {volume} {47}},\ \bibinfo
  {pages} {56} (\bibinfo {year} {2021})}\BibitemShut {NoStop}%
\bibitem [{\citenamefont {Fletcher}\ and\ \citenamefont
  {Reeves}(1964)}]{fletcher1964function}%
  \BibitemOpen
  \bibfield  {author} {\bibinfo {author} {\bibfnamefont {R.}~\bibnamefont
  {Fletcher}}\ and\ \bibinfo {author} {\bibfnamefont {C.~M.}\ \bibnamefont
  {Reeves}},\ }\href {\doibase 10.1093/comjnl/7.2.149} {\bibfield  {journal}
  {\bibinfo  {journal} {Comput. J.}\ }\textbf {\bibinfo {volume} {7}},\
  \bibinfo {pages} {149} (\bibinfo {year} {1964})}\BibitemShut {NoStop}%
\bibitem [{\citenamefont {Sato}(2021)}]{sato2021riemannian}%
  \BibitemOpen
  \bibfield  {author} {\bibinfo {author} {\bibfnamefont {H.}~\bibnamefont
  {Sato}},\ }\href@noop {} {\emph {\bibinfo {title} {Riemannian optimization
  and its applications}}}\ (\bibinfo  {publisher} {Springer},\ \bibinfo {year}
  {2021})\BibitemShut {NoStop}%
\bibitem [{\citenamefont {Pfeuty}(1970)}]{PFEUTY197079}%
  \BibitemOpen
  \bibfield  {author} {\bibinfo {author} {\bibfnamefont {P.}~\bibnamefont
  {Pfeuty}},\ }\href {\doibase https://doi.org/10.1016/0003-4916(70)90270-8}
  {\bibfield  {journal} {\bibinfo  {journal} {Ann. Phys. (N.Y.)}\ }\textbf
  {\bibinfo {volume} {57}},\ \bibinfo {pages} {79} (\bibinfo {year}
  {1970})}\BibitemShut {NoStop}%
\bibitem [{\citenamefont {Chepiga}\ and\ \citenamefont
  {Mila}(2017)}]{PhysRevB.96.054425}%
  \BibitemOpen
  \bibfield  {author} {\bibinfo {author} {\bibfnamefont {N.}~\bibnamefont
  {Chepiga}}\ and\ \bibinfo {author} {\bibfnamefont {F.}~\bibnamefont {Mila}},\
  }\href {\doibase 10.1103/PhysRevB.96.054425} {\bibfield  {journal} {\bibinfo
  {journal} {Phys. Rev. B}\ }\textbf {\bibinfo {volume} {96}},\ \bibinfo
  {pages} {054425} (\bibinfo {year} {2017})}\BibitemShut {NoStop}%
\end{thebibliography}

\end{document}